\documentclass{amsart}
\usepackage[scale=0.81]{geometry}                		
\usepackage{graphicx}				
\usepackage{amssymb,amsmath,color}
\usepackage{xspace}

\providecommand{\keywords}[1]{\textbf{\textit{Keywords---}} #1}
\DeclareMathAlphabet\mathbfcal{OMS}{cmsy}{b}{n}
\usepackage{hyperref}
\usepackage{natbib}
\hypersetup{
    citecolor=black,
    filecolor=black,
    linkcolor=black,
    urlcolor=black
}

\usepackage[T1]{fontenc}
\usepackage[utf8]{inputenc}
\usepackage{float}
\usepackage{setspace}
\usepackage{amsfonts,amsthm,mathrsfs}
\usepackage{psfrag,epsf}
\usepackage{enumerate}
\usepackage{url} 

 \usepackage[french,ruled]{algorithm2e}
\newenvironment{procedureBLLiM}[1][htb]
  {\renewcommand{\algorithmcfname}{BLLiM procedure}
   \begin{algorithm}[#1]%
  }{\end{algorithm}}

  \makeatletter
\newcommand{\RemoveAlgoNumber}{\renewcommand{\fnum@algocf}{\AlCapSty{\AlCapFnt\algorithmcfname}}}
\newcommand{\RevertAlgoNumber}{\algocf@resetfnum}

\usepackage[english]{babel}

\def\XS{\xspace}
\DeclareMathAlphabet{\mathb}{OML}{cmm}{b}{it}
\def\sbm#1{\ensuremath{\mathb{#1}}}                
\def\sbmm#1{\ensuremath{\boldsymbol{#1}}}          

\def\Ab{{\sbm{A}}\XS}  
  \def\bb{{\sbm{b}}\XS}
\def\Cb{{\sbm{C}}\XS}  \def\cb{{\sbm{c}}\XS}

  \def\fb{{\sbm{f}}\XS}
  \def\gb{{\sbm{g}}\XS}
  \def\hb{{\sbm{h}}\XS}

\def\mb{{\sbm{m}}\XS}

\def\Vb{{\sbm{V}}\XS}

\def\Gammab    {{\sbmm{\Gamma}}\XS}

\def\thetab      {{\sbmm{\theta}}\XS}

      \def\Sigmab    {{\sbmm{\Sigma}}\XS}
  
        \def\Phib       {{\sbmm{\Phi}}\XS}

\def\PM{\kern0pt^{\textrm{{\scriptsize PM}}}\kern0pt}
\def\MMAP{\kern1pt^{\textrm{{\tiny MMAP}}}\kern-1pt}

\def\Prob{\mathbb{P}} 
\def\Var{\text{Var}}
\def\Tr{\text{Tr}}

\newcommand{\wv}{\ensuremath{\mathbf{w}}} 
\newcommand{\Xv}{\ensuremath{\mathbf{X}}} 
\newcommand{\xv}{\ensuremath{\mathbf{x}}} 
\newcommand{\Yv}{\ensuremath{\mathbf{Y}}} 
\newcommand{\yv}{\ensuremath{\mathbf{y}}} 

\def\wp1{\mathrm{w.p.} 1}  

\usepackage{amsmath}

\begin{document}

\title{
Nonlinear network-based quantitative trait prediction from transcriptomic data 
 }


%
\author{Emilie Devijver}\address{Department of Mathematics and Leuven Statistics Research Center (LStat), KU Leuven, Leuven, Belgium}
\author{M\'elina Gallopin}\address{Institut de Biologie Int\'egrative de la Cellule, Universit\'e Paris-Sud, 91405 Orsay cedex, France}
\author{Emeline Perthame}\address{ INRIA Grenoble Rh\^one-Alpes, 655 avenue de l'Europe, 38330 Montbonnot Saint-Martin, France}

%
%
%
\maketitle

\begin{abstract}

Quantitatively predicting phenotypic variables by the expression changes in a set of candidate genes is of great interest in molecular biology but it is also a challenging task for several reasons. First, the collected biological observations might be heterogeneous and correspond to different biological mechanisms. Secondly, the gene expression variables used to predict the phenotype are potentially highly correlated since genes interact through unknown regulatory networks. In this paper, we present a novel approach designed to predict quantitative traits from transcriptomic data, taking into account the heterogeneity in biological observations and the hidden gene regulatory networks. The proposed model performs well on prediction but it is also fully parametric, which facilitates the downstream biological interpretation. The model provides clusters of individuals based on the relation between gene expression data and the phenotype, and also leads to infer a gene regulatory network specific for each cluster of individuals.
We perform numerical simulations to demonstrate that our model is competitive with other prediction models, and we demonstrate the predictive performance and the interpretability of our model to predict olfactory behaviour from transcriptomic data on real data from Drosophila Melanogaster Genetic Reference Panel  (DGRP).

\end{abstract}

\section{Introduction}
The development of high-throughput technologies enables to gain insight into complex biological mechanisms at the genome, proteome or transcriptome level, and explain 
biological variability of phenotypes. The analysis of "omic" data is a challenging task and extensive efforts have been made to provide a large range of methods to 
extract information from the data. For transcriptomic data, a large range of original methods has been proposed: co-expression analysis \cite{Eisen1998}, 
phenotype prediction \cite{Golub1999}, gene regulatory network inference \cite{Butte2000}, continuous phenotype prediction \cite{Datta2001}, observations clustering 
\cite{Nguyen2002} or differential analysis \cite{Smyth2004}. 
New improvements propose to combine different techniques. In particular, advances have been made to study complex networks \cite{Barabasi1999} and have appeared to be relevant in biology \cite{Barabasi2004,Fu2013}. Classical approaches are turned into network-based approaches to take into account known network-structured prior information extracted from databases (for instance, from Kyoto Encyclopedia of Genes and Genomes (KEGG) \href{http://www.genome.jp/kegg}{http://www.genome.jp/kegg}), or network-structured information inferred directly from the data. Network information collected on external knowledge databases have been used to improve differential analysis \cite{Jacob2012} or sample classification \cite{Chuang2007}. \cite{Valcarcel2011} and \cite{DelaFuente2010} have proposed differential network analysis. \cite{Valcarcel2014} have proposed to combine metabolic network analysis and genome-wide association studies. Convergence of different methods of analysis is the key to unravel the complexity of biological mechanism and extract more information from the data. 

\smallskip

In this paper, we focus on the problem of predicting a small set of continuous quantitative traits from gene expression profiles. The predicted variables can 
be any continuous measurements, such as the abundance of several proteins, the level of expression of several genes, body mass index, drug tolerance, response index to stress conditions or other complex 
traits. The variables used to predict the phenotype are gene expression data measured by microarrays or by the RNA-seq technologies. To perform the prediction of the phenotype, we 
propose a fully parametric model combining sample clustering and gene regulatory network inference. A clustering of the observations 
allows to catch the nonlinear relationship between the variables to predict and the expression data. Gene network inference for each cluster of observations 
allows to detect modules of regulated genes and improve phenotype prediction. We illustrate our method on the prediction of the response index to benzaldehyde measured in \cite{Swarup2013} from gene expression data measured in \cite{Huang2015}. Both datasets have been generated independently, for drosophila sharing the same genotype in \cite{Swarup2013} and \cite{Huang2015} respectively, in the context of the Drosophila Melanogaster Reference Panel Project  (DGRP) \cite{Mackay2012}. 

\smallskip

The use of gene expression data to predict phenotypic variables is not new: many methods have used transcriptomic data to classify observations or predict 
phenotypic states, as for example disease versus non disease or tumor versus normal \cite{Golub1999, Chuang2007, Nguyen2002}. In this paper, we focus on continuous phenotypical variables  (i.e. quantitative traits) 
prediction from expression data. For example, \cite{Datta2001} has predicted the level of expression of a given gene, \cite{Yang2011} have predicted the 
blood level of several plasma metabolites, \cite{Mach2013} have predicted complex immune traits and \cite{Suzuki2014} have predicted antibiotic resistance. 
The models used to perform such predictions are based on linear regressions. They mainly focus on variable selection to reduce the high-dimension of the problem, 
such as sparse partial least square regression \cite{LeCao2008} or prior genetic algorithm to select the variables \cite{Suzuki2014}. In our case, we assume that
we have already selected relevant genes. Our goal is to improve the prediction while keeping 
a fully parametric and interpretable model. Developing interpretable model, easily understandable by human experts, is relevant in prediction as illustrated by \cite{Letham2015}. 
For that purpose, we use an inverse regression approach, initially introduced by \cite{L91} through the Sliced Inverse Regression method, usually denoted SIR. Several extensions of SIR have afterwards been proposed: see for example Kernel SIR for a nonlinear extension of SIR \cite{W08}, Robust SIR \cite{DYZ15} or Regularized SIR \cite{ZZ05} for a high-dimensional extension of this method. In this paper, we choose the inverse regression approach described in \cite{DFBH15,DFH14,PFD16}. 

\smallskip
The methods cited to perform continuous phenotype  (linear regression or PLS regression) assume that there is a linear relationship between the 
variable to predict and the gene expression data. However, the link function relating the quantitative trait to explanatory variables is potentially 
complex and therefore nonlinear \cite{Torres-Garcia2009}. Non model-based approaches are very efficient to deal with real datasets and nonlinearity. Some examples of non model-based approaches are support vector machine  (SVM) \cite{V98}, relevant vector machine \cite{T01} and random forest \cite{B01}. In contrast, model-based approaches are easier to interpret. However, several nonlinear approaches, based on least squares, require to specify the form of the link function as in the generalized linear model. The prior information about the underlying link function is not always available and misspecification of the link function may lead to poor predictive performances. To avoid this issue, we assume that the model to predict the phenotypic variables from the transcriptomic data is locally  linear. This approach has already been proposed before: we mention the Expectation-Maximization-based Piecewise Regression  (EMPRR) proposed by \cite{AS04}, K-plane proposed by 
\cite{MS15} and the Gaussian locally-linear mapping  (GLLiM) introduced by \cite{DFH14}. 
This assumption 
implies to detect sets of individuals in the sample for which the relationship between the phenotypic variables and the gene expression data is linear. The 
main advantage of this mixture of linear regressions approach is the interpretation given to inferred clusters of individuals, which may be biologically meaningful, e.g., correspond to different biological 
contexts or different disease states. 

\smallskip
Current methods perform prediction from gene expression data without directly modeling the relationship between genes in the prediction model. 
Partial least square (PLS) regressions take into account the relationship between variables indirectly, using latent factors. We refer to  \cite{HTBbook} for a good introduction on PLS, and to \cite{WKS89,KBT08} for its nonlinear counterpart.
However, latent factors are not easy to interpret from a biological point of view. For transcriptomic data, a large range of methods have been proposed to perform network inference. In \cite{Friedman2008}, the level of expression of each gene is modeled by a Gaussian random variable and the network inference method seeks to retrieve the conditional dependencies between variables. Those conditional dependencies between variables are encoded in the precision matrix of the model, and are interpreted as regulatory relationships between genes. However, estimation of a regularized version of the precision matrix is difficult when the number of observations is low compared to the number of genes. In this context, \cite{DG16} have proposed non-asymptotic model selection tools to detect modules of co-regulated genes from the sample covariance matrix.
 In this article, we generalize the GLLiM procedure to take into account the hidden modules-structured gene regulatory network. The distribution of genes conditionally on the quantitative trait variable to predict is expressed in a closed form in the GLLiM framework \cite{DFH14}. For each cluster of individuals, we detect specific modules of genes based an appropriate model selection tools, following ideas developed in \cite{DG16}. The prediction model we proposed, called Block-diagonal covariance for Gaussian Locally-Linear Mapping  (BLLiM), combines individuals clustering and inference of modules of genes to improve performance prediction and downstream analysis.

\smallskip

This paper is organized as follows. In Section~\ref{sec:modelintro}, the BLLiM model is introduced. The nonlinear prediction model and the proposed estimation procedure based on block decomposition of covariance matrices are described. In Section~\ref{sec:simus}, a 
simulation study demonstrates the prediction accuracy of our method, on data simulated under the model and on data simulated under regression functions harder to estimate, simulated with hidden variables. The prediction accuracy of the proposed method is compared to {state-of-the-art} model-based and non model-based regression methods. In Section~\ref{sec:realdata}, a real data analysis demonstrates that the proposed model achieves accurate prediction while providing interesting interpretation tools both on clusters of observations and inferred gene regulatory networks. This paper ends with a discussion in Section~\ref{sec:conclu}. Details on the estimation algorithm are relegated to an Appendix.

\section{A network-based prediction model\label{sec:modelintro}}

In this paper, we propose a model to perform the prediction of a multivariate response $\Yv \in \mathbb{R}^L$ from a set of covariates $\Xv =  (X_1,\ldots,X_D)$. On the real dataset studied in Section \ref{RealData}, $\Yv$ is a matrix of $L$ quantitative traits and $\Xv$ is a matrix of expression for $D$ genes.
First of all, we consider a model-based clustering which allows to account for heterogeneous data, and to predict the response through a nonlinear model.
To deal with high-dimensional data, an inverse regression trick is used, 
performing first the regression of $\Xv$ on $\Yv$, and 
inverting the system of equations. Due to the correlations between genes, we consider a non diagonal covariance matrix, but focus on sparse structure to tackle the 
dimension issue. 
Moreover, we focus on the strength of the model proposed in this paper to interpret parameters: tools are provided to understand 
the relation between covariates and response.
 Then, an interpretation is obtained within each cluster. 
 

In this section, we describe the model, the estimation procedure, and the algorithm implementing the method. 

\subsection{Modelisation}
\subsubsection{Finite mixture of regressions model for prediction\label{sec:model}}
\label{model}
We propose a general model-based procedure to predict a multivariate response $\Yv$ from a set of covariates $\Xv$. 
To catch the nonlinear relationship between the response and the set of covariates, we propose to approximate the regression function of interest by $K$ locally affine regression functions considered as several clusters. The latent variable $Z$ is introduced to describe the cluster membership: $Z_i=k$ if the individual $i\in\{1,\ldots, n\}$ originates from cluster $k \in \{1, \ldots, K \}$.
The joint distribution of $ (\Yv,\Xv)$ 
 is considered to be a finite mixture of multivariate Gaussian distributions: for the individual $i \in \{1,\ldots, n\}$,
\begin{eqnarray}
p (\Yv_i=\yv, \Xv_i=\xv | Z_i=k) = \varphi_{L+D} ([\yv,\xv]^T;\mb_k^*, \Vb_k^*), \label{def:joint} 
\end{eqnarray}
where $\varphi_{L+D}$ denotes the density function of a Gaussian distribution of dimension $L+D$, and $\mb_k^*$ and $\Vb_k^*$ respectively the mean and variance parameters of the Gaussian distribution. 

Without loss of generality, we decompose $\mb_k^*$ and $\Vb_k^*$ to reparametrize the model:
\begin{equation}
 \begin{array}{rl}
 \mb_k^*=&\left[
 \begin{array}{c}
 \Ab_k^*\cb_k^*+\bb^*_k \\
 \cb_k^*
 \end{array}
 \right]; \\
 \Vb_k^*=&\left[
 \begin{array}{cc}
 \Sigmab_k^*+\Ab_k^*\Gammab_k^* (\Ab_k^*)^T & \Ab_k^*\Gammab_k^* \\
 \Gammab_k^* (\Ab_k^*)^T & \Gammab_k^*
 \end{array}
 \right]. \label{eq:JGMMtoGLM}
 \end{array}
\end{equation}
This reparameterization allows to consider the conditional distributions with the following notations, for a given individual $i\in \{1,\ldots, n\}$ coming from cluster $k \in \{1,\ldots, K\}$:
\begin{align}
& p (\Xv_i=\xv | Z_i=k) = \varphi_{D} (\xv ; \cb^*_k,\Gammab^*_k);
 \label{eq:Xznou2} \\
 & p (\Yv_i=\yv|\Xv_i=\xv, Z_i=k) 
= \varphi_{L} (\yv ;\Ab^*_k \xv + \bb^*_k 
,\Sigmab^*_k);
 \label{eq:Yxznou2}
\end{align}
where $\cb^*_k$ and $\Gammab^*_k$ characterize the distribution of the covariates in cluster $k$, independently of the response $\Yv_i$, and where $\Ab_k^*$ are the coefficients 
of the linear regression of $\Yv_i$ on $\Xv_i$ if the individual $i$ belongs to the cluster $k$, and $\Sigmab_k^*$ is the residual covariance matrix 
 in the corresponding regression.

For prediction, by integrating Equations~\eqref{eq:Xznou2} and \eqref{eq:Yxznou2} according to the latent variable $Z$, it appears that the conditional 
distributions are weighted multivariate Gaussians, for a given individual $i \in \{1,\ldots,n\}$:
\begin{align}
& p (\Yv_i=\yv|\Xv_i=\xv) = 
 \label{eq:JGMM_inverse_map}
\sum_{k=1}^K\frac{\pi_k^ * \varphi_D (\xv ; \cb_k^*,\Gammab_k^*)}{\sum_{j=1}^K \pi_j^* \varphi_D (\xv ; \cb_j^*,\Gammab^*_j)}
 \varphi_L (\yv;\Ab^*_k\xv+\bb^*_k
 ,\Sigmab_k^*).
\end{align}
In the above, $\pi_k^*$ denotes the proportion for class $k \in \{1,\ldots, K\}$, such that $\pi_1^* + \ldots \pi_K^* =1$. We denote by $\thetab^*_K$ the vector of parameters:
$$ \thetab^*_K =  (\cb_k^*,\Gammab_k^*, \Ab_k^*,\bb_k^*, \Sigmab_k^*)_{1\leq k \leq K}\in \Theta_K^* =  (\mathbb{R}^D\times \mathcal{S}_D^{++} (\mathbb{R}) \times \mathbb{R}^{L\times D}\times \mathbb{R}^L\times \mathcal{S}_L^{++} (\mathbb{R}))^K;$$
where $\mathcal{S}_D^{++} (\mathbb{R})$ denotes the collection of symmetric positive definite matrices on $\mathbb{R}^D$.
A prediction $\hat{\Yv}_{n+1}$ of the response $\Yv_{n+1}$ from a new vector of covariates $\Xv_{n+1}$ is achieved by taking the expectation in Equation~\eqref{eq:JGMM_inverse_map}. 
Indeed, the probability of a new observation of covariates $\xv_{n+1}$ to belong 
to each cluster 
is estimated, and the prediction is computed afterwards by a linear combination of the linear models associated to each cluster such as we have:
\begin{align*}
\hat \Yv_{n+1} &= \mathbb E (\Yv_{n+1}|\Xv_{n+1}=\xv_{n+1}) = 
\sum_{k=1}^K\frac{\pi_k^ * \varphi_D (\xv_{n+1} ; \cb_k^*,\Gammab_k^*)}{\sum_{j=1}^K \pi_j^* \varphi_D (\xv_{n+1} ; \cb_j^*,\Gammab^*_j)}
\left (\Ab^*_k\xv_{n+1}+\bb^*_k \right),
\end{align*}
where $\hat \Yv_{n+1}$ denotes the prediction of a new response by the model.

\subsubsection{Inverse regression model}
\label{IR}

There are many covariates compared to potentially low number of individuals in each cluster, and too much parameters  (detailed in $ \thetab^*_K$) have to be estimated in the linear model associated to each cluster. 
The number of parameters to estimate is reduced by using the inverse regression trick of \cite{DFH14} in conjunction with a block-diagonal structure hypothesis on the residual covariance matrices $\Sigmab_k$. 
Note that making an assumption on residual covariance matrices $\Sigmab_k$ is more realistic than making an assumption on covariance matrices of covariates $\Gammab_k^*$, which justifies the use of the inverse regression trick to reduce the dimension.  

Coming back to Equation \eqref{def:joint}, we first consider $\Yv$ as the covariates, and $\Xv$ as the multivariate response.
Then, the conditional distribution, called the \emph{forward conditional distribution function}, is defined by, for an individual $i \in \{1,\ldots,n\}$ coming from cluster $k \in \{1,\ldots, K\}$:
\begin{align}
 & p (\Yv_i=\yv | Z_i=k) = \varphi_{L} (\yv ; \cb_k,\Gammab_k);
 \label{eq:Xznou} \\
&
 p (\Xv_i=\xv | \Yv_i=\yv, Z_i=k) 
= \varphi_{D} (\xv ;\Ab_k \yv + \bb_k
,\Sigmab_k);
 \label{eq:Yxznou}
\end{align}
where $ (\cb_k,\Gammab_k,\Ab_k,\bb_k,\Sigmab_k)$ are deduced from $ (\cb_k^*,\Gammab_k^*,\Ab_k^*,\bb_k^*,\Sigmab_k^*)$.
The following one-to-one correspondence defines the link between the inverse conditional distribution and the forward conditional distribution function:
\begin{align*}
\Psi: & \Theta_K \rightarrow \Theta^*_K\\
&\thetab_K \mapsto \thetab^*_K\\
&\begin{pmatrix}
 \cb_k\\
 \Gammab_k \\
 \Ab_k \\
 \bb_k \\
 \Sigmab_k
 \end{pmatrix}_{1\leq k \leq K}
\mapsto
\begin{pmatrix}
 \cb_k^*\\
 \Gammab_k^* \\
 \Ab^*_k \\
 \bb^*_k \\
 \Sigmab_k^*
 \end{pmatrix}_{1\leq k \leq K}
 =
 \begin{pmatrix}
 \Ab_k\cb_k\\
 \Sigmab_k+\Ab_k\Gammab_k\Ab_k^T\\
 \Sigmab_k^*\Ab_k^T\Sigmab_k^{-1}\\
 \Sigmab_k^* (\Gammab_k^{-1}\cb_k-\Ab_k^T\Sigmab_k^{-1}\bb_k)\\
  (\Gammab_k^{-1}+\Ab_k^T\Sigmab_k^{-1}\Ab_k)^{-1}
 \end{pmatrix}_{1\leq k \leq K}.
\end{align*}

Without assuming anything on the structure of parameters, this model counts
$$K\left (L + \frac{L (L+1)}{2} + D (L+1) + \frac{D (D+1)}{2} +1\right)-1$$
parameters to estimate, which is large when $D$ is large. For example, in the real dataset we analyze in Section~\ref{sec:realdata}, we have $L=1$, $D=50$, and the size of the sample is $n=326$. The selected number of affine components is equal to $K=3$. The total number of parameters is larger than $4~000$ if we consider full residual covariance matrices $ (\Sigmab_k)_{1\leq k\leq K}$ in equation \eqref{eq:Yxznou}. If we consider that the residual covariance matrices $ (\Sigmab_k)_{1\leq k\leq K}$ are diagonal, as assumed in \cite{DFH14}, the number of parameters goes down to less than $500$. However, this assumption is unrealistic in the context of transcriptomic data: the matrices $ (\Sigmab_k)_{1\leq k\leq K}$ correspond to matrices of correlations between covariates  (in our real data, genes) conditionally on the response  (in our real data, the response index to benzaldehyde). 

\subsubsection{Block diagonal covariance matrices}

Genes are known to interact through an unknown regulatory network associated with the phenotypic response. 
 {To make a trade-off between complexity and sparsity, block-diagonal covariance matrices $ (\Sigmab_k)_{1\leq k\leq K}$ are considered, up to a permutation of covariates. It boils down to assume that, conditionally on the phenotypic response, genes interact with few other genes only, i.e. there are small modules of correlated genes, in the spirit of the work developed in \cite{DG16}. Note that the matrix $\Sigmab_k$ corresponds to the residual covariance of the covariates $\Xv$ conditionally on $\Yv$ for the cluster $k \in \{1,\ldots, K\}$. }

The notations used to index the groups of correlated covariates are introduced here: first, remark that each set of groups is specific to each cluster of individuals. For a given cluster $k \in \{1,\ldots, K\}$, we decompose $\Sigmab_k$ into $G_k$ blocks, 
and we denote by $d_k^{[g]}$ the set of variables into the $g$th group, for $g \in \{1,\ldots,G_k\}$, and by $\#\{d_k^{[g]}\}$ the number of variables in the corresponding set.
Up to a permutation, we may write the covariance matrices as a block diagonal covariance matrix: for $k \in \{1,\ldots, K\}$, if $B_k$ defines the blocks, for the cluster $k$,
\begin{align}
\Sigmab_k (B_k) &= P_k
\begin{pmatrix}
\Sigmab_k^{[1]} & 0 & \ldots & 0 \\
0 & \Sigmab_k^{[2]} &\ldots & 0\\
0&0&\ddots & 0\\
0&0&\ddots &\Sigmab_k^{[G_k]} 
\end{pmatrix} P_k^{-1} ;
 \label{SigmaBlock}
\end{align}
where $P_k$ corresponds to the permutation matrix in cluster $k$, and $\Sigmab_k^{[g]} \in \mathcal{S}_{\#\{d_k^{[g]}\}}^{++} (\mathbb{R})$ corresponds to the residual correlations between 
the $\#\{d_k^{[g]}\}$ variables in group $g \in \{1,\ldots, G_k\}$.

With this reduction, the number of parameters to estimate $ \Delta_{ (K,\mathbf{B})}$ is equal to
\begin{align}
 \Delta_{ (K,\mathbf{B})} &= K\left (L + \frac{L (L+1)}{2} + D (L+1) +1 \right) + \sum_{k=1}^K \sum_{g=1}^{G_k} \frac{\#\{d_k^{[g]}\} (\#\{d_k^{[g]}\}+1)}{2}- 1
 \label{DeltaKB}
\end{align}
which is smaller than previously if the blocks are small enough. Following the real example from Section~\ref{sec:realdata}, based on the block diagonal structure, the number of parameters falls to $1~216$, which is slightly more than the number of parameters under an heterotropic assumption but still far from the number of parameters of full matrix. 

Finally, the estimation procedure consists in first estimating 
$$\thetab_K (\mathbf{B}) =  (\cb_k,\Gammab_k,\Ab_k,\bb_k,\Sigmab_k (B_k))_{1\leq k \leq K} \in \Theta_K =  (\mathbb{R}^L\times \mathcal{S}_L^{++} (\mathbb{R}) \times \mathbb{R}^{D\times L}\times \mathbb{R}^D\times \mathcal{S}_D^{++} (\mathbb{R}))^K,$$ 
with $\Sigmab_k (B_k)$ structured in blocks for all $k \in \{1,\ldots, K\}$, 
and to deduce $\thetab^*_K =  (\cb_k^*,\Gammab_k^*,\Ab_k^*,\bb_k^*,\Sigmab_k^*)_{1\leq k \leq K}$.

Remark that for any $k \in \{1,\ldots, K\}$, the block-diagonal covariance structure $B_k$ of $\Sigmab_k (B_k)$ leads to a decomposition of $\Gammab^*_k$ into a sum of a block diagonal 
matrix $\Sigmab_k$ and a low rank matrix described by $\Ab_k\Gammab_k^{1/2}$. 
Without any assumption on the sparse and low rank structures, this decomposition is ill-posed and intractable. Many authors proposed methods based on convex optimization to disentangle the sparse structure from the low rank part. 
For example, \cite{lowrank11} assume some hypothesis on the space of row and column of the low rank matrix and on the distribution of sparsity in the sparse matrix to make the problem identifiable. On the other hand, \cite{robustPCA} suppose that sparsity is uniform in the sparse matrix and make some assumptions about the incoherence of the low rank part, 
showing that under these minimal assumptions, it is possible to estimate such a decomposition.
In this work, the block-diagonal structure is automatically detected and 
identifiability is insured by the inverse regression trick, as both the sparse and low rank parts are estimated by performing standard linear regression estimation  (see Section~\ref{sec:algo} for more details). 

\subsection{BLLiM procedure}\label{sec:sel}
\label{SlopeHeuristics}
In this paper, 
 individuals are clustered into homogeneous clusters,
and covariates are organized into independent groups. 
 To avoid confusion, the word \textit{clusters} always refers to clusters of individuals, the words \textit{group} and \textit{partition} always refer to groups of variables, or partition of variables. 
The procedure requires to learn the number of clusters and the network structure of covariates, which is performed through a selection of a model among a collection of potential candidates.
\paragraph{Model $F_{ (K,\mathbf{B})}$}
The proposed model is defined by $K$, the number of clusters, and $\mathbf{B} =  ((d_k^{[g]})_{1\leq g \leq G_k})_{1\leq k \leq K}$ the covariate indexes into each group for each cluster.
For a fixed $K$ and a fixed $\mathbf{B}$, we denote by $F_{ (K,\mathbf{B})}$ the corresponding model:
\begin{align*}
 F_{ (K,\mathbf{B})} &= \left\{ \xv \mapsto f_{ (K,\mathbf{B})}  (\xv|\yv) = \sum_{k=1}^K\frac{\pi_k \varphi_L (\yv ; \cb_k,\Gammab_k)}{\sum_{j=1}^K \pi_j \varphi_L (\yv ; \cb_j,\Gammab_j)}
 \varphi_D (\xv;\Ab_k\yv+\bb_k
 ,\Sigmab_k (B_k)) \right\}.
\end{align*}

\paragraph{Collection of models}
The number of clusters $K$ varies among a list of candidate numbers of clusters $\mathcal{K}$ and $\mathbf{B}$ varies among a list of candidate structures $\mathcal{B}$, where $\mathcal{B}$ denotes the set of partitions of the covariables indexed by $\{1,\ldots, D\}$ for each cluster of individuals. We obtain a collection of models
\begin{align*}
 \mathcal{F} &=  (F_{ (K,\mathbf{B})})_{K \in \mathcal{K}, \mathbf{B} \in \mathcal{B}}.
\end{align*}
We upper bound the number of clusters by $K_\text{max}$ and restrict our attention to a finite set $\mathcal{K} \subset \{1,2,\ldots,K_\text{max}\}$.
As the cardinal of $\mathcal{B}$ is large  (Bell number),
 a random subcollection $\mathcal{B}^R$ of moderate size is considered, using the following idea. 

First, 
 the parameters are initialized by estimating Model~\eqref{eq:JGMM_inverse_map} with diagonal constraint on $ (\Sigmab_k)_{1\leq k \leq K}$ using GLLiM \cite{DFH14}. 
From \eqref{eq:Yxznou}, for each $k \in \{1,\ldots,K\}$, a non-diagonal estimator for $\Sigmab_k$ is considered using the decomposition of the variance:
$$\Var (\Xv_i \vert Z_i=k) = \Ab_k \Gammab_k \Ab_k^T +\Sigmab_k \hspace{1cm} \Leftrightarrow
\hspace{1cm} \Sigmab_k = \Var (\Xv_i \vert Z_i=k) - \Ab_k \Gammab_k \Ab_k^T .$$
$\hat{\Ab}_k$ and $\hat{\Gammab}_k$ are available from the initialization step, whereas $\Var (\Xv_i \vert Z_i=k)$ is estimated using the sample covariance matrix of individuals belonging to cluster $k$. 
For each $k \in \{1,\ldots,K\}$, a collection of block diagonal covariance matrices is then built by thresholding $\hat{\Sigmab}_k$ at level $\lambda$:
 \begin{align*}
 [\hat{\Sigmab}_k^{[\lambda]}]_{j_1,j_2}= \left\{ 
	\begin{array}{ll}
 0 & \text{ if } \vert [\hat{\Sigmab}_k]_{j_1,j_2} \vert \leq \lambda; \\ 
 \phantom{ (} \hspace{-0.3cm} [\hat{\Sigmab}_k]_{j_1,j_2} & \text{ elsewhere,}
 \end{array}
 \right.
\end{align*}
where $|.|$ denotes the absolute value.
Potential thresholds are described by $\{|[\Sigmab_k]_{j_1,j_2}|\}_{1\leq j_1 \leq j_2 \leq D}$. 
{For each $k \in \{1,\ldots, K\}$, the structure of $\Sigmab_k$ is summarized by its block diagonal structure $B_k (\lambda)$, restricting the collection to at most D(D+1)/2 models corresponding to thresholds denoted by $ (\lambda_k^{[j]})_{1\leq j \leq D(D+1)/2}$} {and sorted in increasing order}. Subsequently, we consider the collection of  (potentially different between clusters) block-diagonal structures $ (B_1 (\lambda_1^{[j]}),\ldots, B_K (\lambda_K^{[j]}))_{1 \leq j \leq D(D+1)/2}$, with the constraint that for a fixed $j \in \{1,\ldots, D(D+1)/2\}$, every covariance matrix has a comparable level of sparsity within the clusters. 

 Remark that constructing the collection of models in the initialization step allows to approximate the block-diagonal structure without estimating the model collection at each step of the EM 
algorithm, which drastically reduces the computation time.

\paragraph{Model selection}
Considering the collection of models
\begin{align*}
 \mathcal{F}^R &=  (F_{ (K,\mathbf{B})})_{K \in \mathcal{K}, \mathbf{B} \in \mathcal{B}^R},
\end{align*}
we select a model using the slope heuristic introduced in \cite{BirgeMassart} and described here. First, if at least two models have the same dimension, the one which maximizes the log-likelihood is kept, the dimension of a model being defined by the number of parameters we have to estimate, introduced in \eqref{DeltaKB}. 

The selected model is
\begin{align} 
 (\hat K,\hat{\mathbf{B}}) = \underset{ (K, \mathbf{B})}{\operatorname{argmin}}\left\{ -\frac{1}{n} \sum_{i=1}^n \log (f_{ (K, \mathbf{B})}  (\xv_i|\yv_i)) 
+ \kappa \Delta_{ (K,\mathbf{B})} \right\};
\label{SlopeHeuristic}
\end{align}

where $\kappa$ is learned from the data using the R package \verb?capushe? described in \cite{Baudry}.

The heuristic used to determine $\kappa$ has been proved in very specific cases, as heteroscedastic regression with fixed design 
 \cite{BirgeMassart, Baraud} or homoscedastic regression with fixed design \cite{Arlot2010}. A justification of the penalty shape has also been provided, e.g., in \cite{Devijver}
 for the problem of selecting the number of components in mixture regression, and in \cite{DG16} for selecting the block diagonal structure of covariance matrix. Although we do not provide such a justification in our context, remark that our model is related to the models described in \cite{Devijver} and in \cite{DG16}.
 Moreover, the slope heuristic has already been used without theoretical justification and is known to perform well in several context, for instance, to select the number of components in discriminative functional mixture models \cite{Bouveyron}.
 
Here, a slightly generalization of this method is used by decomposing the problem of minimization \eqref{SlopeHeuristic} into several steps:
first, for each $K \in \mathcal{K}$, we select a set of partitions $\hat{\mathbf{B}}_K$:
\begin{align*}
\hat{\mathbf{B}}_K = \underset{\mathbf{B}}{\operatorname{argmin}}\left\{ -\frac{1}{n} \sum_{i=1}^n \log (f_{ (K, \mathbf{B})}  (\xv_i|\yv_i)) 
+ \kappa_B \Delta_{ (K,\mathbf{B})} \right\}.
\end{align*}
Then, among the models $ (F_{ (K,\hat{\mathbf{B}}_K)})_{K \in \mathcal{K}}$, we select one with the same process:
\begin{align*}
\hat K = \underset{K}{\operatorname{argmin}}\left\{ -\frac{1}{n} \sum_{i=1}^n \log (f_{ (K, \hat{\mathbf{B}}_K)}  (\xv_i|\yv_i)) 
+ \kappa_K \Delta_{ (K,\hat{\mathbf{B}}_K)} \right\}.
\end{align*}
The coefficients $\kappa_{\mathbf{B}}$ and $\kappa_K$ may be different. Remark that this nested slope heuristic is based on non-asymptotic model selection arguments, which are well-suited under high-dimensional contexts.

\paragraph{BLLiM procedure}
The BLLiM procedure is summarized bellow:

\RemoveAlgoNumber
\begin{procedureBLLiM}[H]
\SetAlgoCaptionSeparator{}
\caption{}
{Fix $\mathcal{K}$ a range of candidate number of clusters. For each $K \in \mathcal{K}$: } 
\begin{enumerate}
\item[Step A] Initialization of parameters.\\
Run GLLiM to initialize the estimation of parameters and the clustering.
\item[Step B] Construction of partitions of $ (\Sigmab_k)_{1\leq k\leq K}$.
\begin{enumerate}
\item[ (1)] For all $k\in \{1,\ldots, K\}$, compute $\Sigmab_k = \Var (\Xv_i \vert Z_i=k) - \Ab_k \Gammab_k \Ab_k^T $.
\item[ (2)] For all $k\in \{1,\ldots, K\}$, threshold $\Sigmab_k$ and deduce a collection of partitions related to several block-diagonal structures.
\item[ (3)] Construct a collection of at most $D$ models $\mathcal{B}^R$, with block-diagonal structures for each cluster of individuals, each collection of partition is sorted by sparsity level.
\end{enumerate}
\item[Step C] Estimation in each model.\\ 
Fix $\mathbf{B} \in \mathcal{B}^R$. Run an EM algorithm to approximate the maximum likelihood estimators. 
\item[Step D] Selection of $\hat{\mathbf{B}}$ using the slope heuristic.
\end{enumerate}
Varying $K\in \mathcal{K}$, we get a collection of models.
Use the slope heuristic to select $\hat{K}$.
\end{procedureBLLiM}

\subsection{EM algorithm\label{sec:algo}}

The framework introduced above has the advantage that estimation of parameters $\thetab_K (\mathbf{B})= (\cb_k,\Gammab_k,\Ab_k,\bb_k,\Sigmab_k (B_k))_{1\leq k \leq K}$ is tractable by an Expectation-Maximization algorithm  (so called EM algorithm). 

%

For each block diagonal structure described by $\mathbf{B}$ and with $K$ fixed, we estimate the parameters with an EM algorithm, described in the following. Introduced in \cite{Dempster}, an EM algorithm consists in alternating two steps until convergence in order to reach a maximum of the likelihood. The outlines of the algorithm are described hereafter and computational details are given in Appendix.
\begin{itemize}
\item E-step\\
The update of individual probabilities 
is computed from the joint distribution function of the observations. This step is classical in the estimation of a joint Gaussian mixture on both responses and covariates. 
\item M-step \\
In the following, the estimators of each parameter are interpreted in terms of statistical estimation problems  (estimation of a Gaussian mixture model  (GMM), of regression coefficients and estimation of the blocks in a block diagonal matrix) weighted by the probability of each observation to belong to cluster $k$. 
For each $k \in \{1,\ldots, K\}$, using the weights deduced in the E-step, we have
\begin{itemize}
\item $\pi_k$, $\cb_k$ and $\Gammab_k$ are estimated by GMM-like estimators as in \cite{DFH14}.
Estimator of $\pi_k$ can be interpreted as the probability of an observation to belong to each class, independently of its profile of response and covariate.
Estimator of $\cb_k$ can be interpreted as the sample mean of $ (\Xv_1,\ldots, \Xv_n)$ weighted by the probability of each observation to belong to cluster $k$. The same remark holds for estimator of $\Gammab_k$ except that it is a reweighed sample covariance matrix of $ (\Xv_1,\ldots, \Xv_n)$.

\item $\Ab_k$ and $\bb_k$ are estimated by Regression-like estimators as in \cite{DFH14}.
Estimator of $\Ab_k$ can be interpreted as the coefficients of a regression of $ (\Xv_1,\ldots, \Xv_n)$ on $ (\Yv_1,\ldots, \Yv_n)$ weighted by the probability of each observation to belong to cluster $k$. Estimator of $\bb_k$ is deduced by estimating the intercept of this regression. 

\item $\Sigmab_k$ is estimated by a block-diagonal estimator. 
For a fixed structure defined by $B_k =  (d_k^{[g]})_{g \in G_k}$, for each $g \in G_k$, the 
sample covariance matrix of the residuals of the regression between $ (\Xv_1, \ldots, \Xv_n)$ and $ (\Yv_1,\ldots, \Yv_n)$ is computed for the restricted set of variables defined by $d_k^{[g]}$.
Then, $\Sigmab_k$ is estimated by 0, except on the blocks defined by $ (d_k^{[g]})_{g \in G_k}$, where we use the mentioned above estimated residual covariance matrix.
\end{itemize}
\end{itemize}

The E and M steps are iterated until the algorithm converges when the growth of log-likelihood is smaller than $10^{-3}$ times the total variation of log-likelihood  (maximum minus minimum of the total progression of the log-likelihood): if $\text{log} \mathbfcal L$ denotes the vector of the log likelihood where $\text{log} \mathbfcal L_l$ is the log-likelihood at step $l$ of the algorithm, the algorithm stops at iteration $l$ if
$$
\text{log} \mathbfcal L_l - \text{log} \mathbfcal L_{l-1} \leq 10^{-3} \left (\max_j (\text{log} \mathbfcal L_j) - \min_j (\text{log} \mathbfcal L_j)\right).
$$

We have included the proposed model and the associated selection procedure in the \texttt{R} package \texttt{xLLiM}, which is freely available from CRAN  (\href{url}{http://cran.r-project.org/web/packages/xLLiM/}).

\section{Simulation study\label{sec:simus}}

In this section, the predictive performances of the proposed model are studied on two distinct simulation settings. In Section~\ref{sec:simu1}, our estimation procedure of the model proposed in Equation~\eqref{eq:JGMM_inverse_map} is validated by comparing the prediction accuracy of the proposed method to other existing versions of the inverse regression approach considered in this paper. Then, in Section~\ref{sec:simus2}, a more intensive simulation study is performed in order to compare BLLiM to model-based and non model-based prediction methods of the literature. 
The code to reproduce the simulation study has been provided as supplementary materials.
\subsection{Prediction of locally affine regression functions\label{sec:simu1}}

\subsubsection{Simulation plan \label{SimulationPlan1}}
\begin{figure}[H]
\begin{center}
\caption{Images of a sample of covariance matrices $ (\Sigmab_1,\ldots,\Sigmab_5)$ for the setting described in Section \ref{SimulationPlan1}. \label{fig:Sigmaa}}
\includegraphics[width=3.3cm]{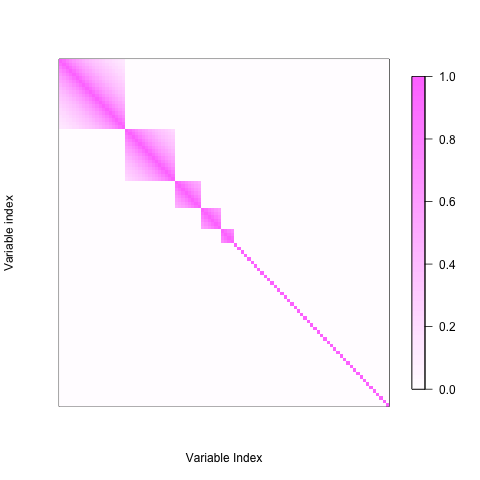}
\includegraphics[width=3.3cm]{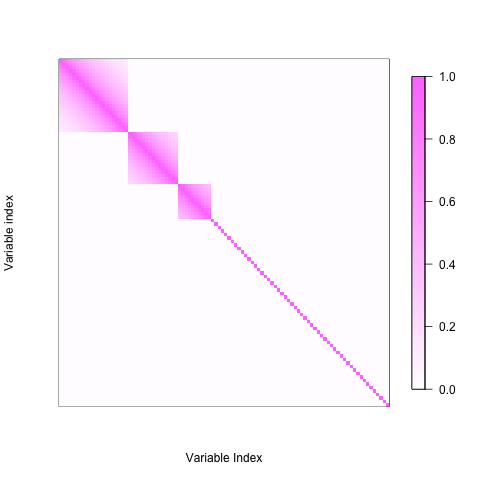}
\includegraphics[width=3.3cm]{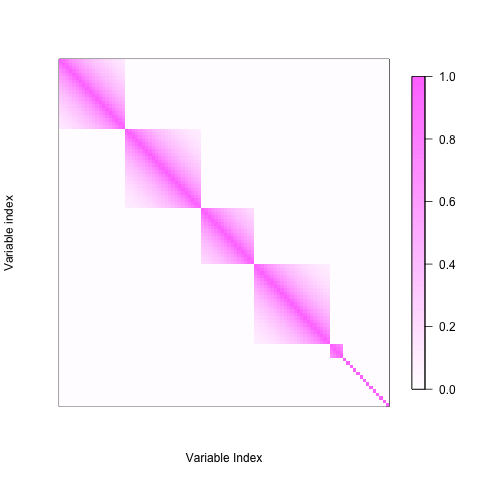}
\includegraphics[width=3.3cm]{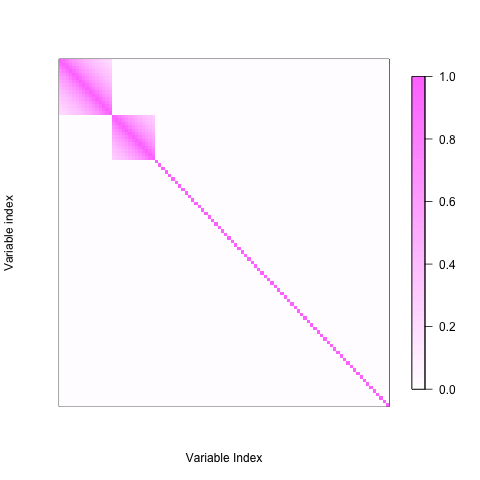}
\includegraphics[width=3.3cm]{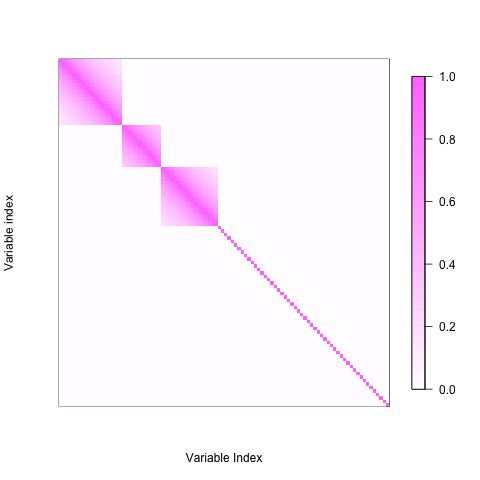}
\end{center}
\end{figure}

The following simulation study aims to demonstrate that BLLiM is able to improve prediction of GLLiM while keeping it numerically tractable. Moreover, in this section, 
the impact of the number of observations on prediction accuracy of the proposed method is assessed. We consider a simulation setting with bivariate response  ($L=2$) and $D=100$ covariates. 
 We consider $100$ replicates generated from model \eqref{eq:JGMM_inverse_map} with $K=5$ components and randomly sampled $\thetab$ as follows:
\begin{itemize}
\item $ (\pi_k)_{1\leq k \leq K}$ are sampled from a uniform $\mathcal U  (0,1)$ and normalized such as $\sum_{k=1}^K \pi_k=1$;
\item $ (\cb_k)_{1\leq k \leq K}$ and $ (\bb_k)_{1\leq k \leq K}$ are sampled from a standard Gaussian and $ (\Ab_k)_{1\leq k \leq K}$ are sampled from a centered Gaussian with standard error set to $\sqrt{0.5}$ in order to reach a SNR  (defined hereafter) around 2 {which corresponds to a reasonably challenging regression problem};
\item $ (\Gammab_k)_{1\leq k \leq K}$ are $L \times L$ correlation matrices generated using the $\texttt{genPositiveDefMat}$ function of the \texttt{clusterGeneration R} package;
\item $ (\Sigmab_k)_{1\leq k \leq K}$ are $D \times D$ block-diagonal matrices. The size and the number of blocks are randomly chosen. We consider the following structure of dependence in the blocks: each block is a Toeplitz matrix with an auto-correlation parameter set to 0.9. In Figure~\ref{fig:Sigmaa}, we represent images of a sample of
covariance matrices 
$ (\Sigmab_k)_{1\leq k \leq K}$ for $K=5$. Note that the number and the size of blocks differ between clusters. 
\end{itemize}

To characterize the difficulty of a simulation setting, we consider the definition of SNR  (Signal-to-Noise Ratio) proposed by \cite{D15} as a multivariate counterpart of standard SNR when parameters are matrices. We define by $\text{SNR}_k$ the signal-to-noise ratio restricted to cluster $k \in \{ 1\ldots, K\}$ defined as: 
$$\text{SNR}_k = \frac{\Tr (\Var (\Yv \vert Z=k))}{\Tr (\Var (\Yv \vert Z=k,\bb_k=0))} = \frac{\Tr (\Ab_k \Gammab_k \Ab_k^T+ \Sigmab_k)}{\Tr (\Sigmab_k)}.$$
The SNR of the whole simulation plan across all clusters is defined by:
$$\text{SNR} = \sum_{k=1}^K \pi_k \text{SNR}_k.$$
For example, in the simulation setting described hereafter, we observe the following value on a given sample SNR$_1 =2.210$, SNR$_2=2.009$, SNR$_3= 2.089$, SNR$_4= 1.976$, SNR$_5= 1.930$ and SNR$=2.037$. \\

We consider two distinct sample sizes:
\begin{enumerate}
\item[\textbf{ (A)}] $n=4162$ equals to the true number of parameters  (block diagonal structure for $ (\Sigmab_k)_{1\leq k \leq K}$).
\item[\textbf{ (B)}] In order to assess the sensibility of the compared methods to the number of observations, we also consider a situation with $n=416$ which is 10 times lower than \textbf{ (A)}. 
\end{enumerate}

The predictive performances of the proposed method are compared to two versions of the model defined in Equations~\eqref{eq:Xznou} and \eqref{eq:Yxznou} as follows: 
\begin{itemize}
\item GLLiM: We denote by GLLiM the version of the model proposed by \cite{DFH14} under heterotropic assumption on the matrices $ (\Sigmab_k)_{1\leq k \leq K}$. In this paper, the authors suppose that each matrix $ (\Sigmab_k)_{1\leq k \leq K}$ is diagonal. This very simple version of the model minimizes the number of parameters to estimate and does not involve hyper parameters to estimate but it relies on a strong assumption on covariance structure. As proposed by \cite{DFH14}, the number of affine components $K$ is determined by minimizing BIC criterion. 
\item GLLiM-$L_w$: The authors of \cite{DFH14} also propose an hybrid version of GLLiM, by completing the observed response with latent factors. They estimate not only diagonal matrices $ (\Sigmab_k)_{1\leq k \leq K}$ but also the loadings associated to these hidden variables, within each clusters. The latent factors can be seen as extra unobserved latent responses in the inverse regression, which leads to increase the rank of matrix $\Ab_k\Gammab_k\Ab_k^T$. The number of factors, namely the rank of the loadings matrices has to be determined. The authors propose to use the BIC criterion to estimate both hyper parameters: number of hidden factors $L_w$ and number of affine components $K$.  We denote this method by GLLiM-$L_w$ in the current paper. 
\end{itemize}

\subsubsection{Assessment of prediction accuracy}
We assess the prediction accuracy of a given prediction procedure by computing the Root Mean Square Error  (RMSE). For $1\leq l \leq L$ if $y_{i,l}$ denotes the $l$-th observed response corresponding to a testing observation $i$, if $\hat y_{i,l}$ denotes the $l$-th response predicted by a given method from the testing profile $\xv_i$, then RMSE for response $l$ is computed as follows:
\begin{align}
\text{RMSE} (l) = \sqrt{ \frac{1}{n} \sum_{i=1}^n  (y_{i,l} - \hat y_{i,l})^2}.
\label{def::MSE}
\end{align}
where $n$ is the number of testing profiles. 
RMSE is a common scale-dependent measure, which is on the same scale as the response. Due to its Savage-Bergman form \cite{Savage}, the associated scoring function is strictly consistent with the mean functional relative to the class of the probability distributions on $\mathbb{R}$ whose second moment is finite. As discussed in \cite{Gneiting2011}, this criterion is relevant for models whose prediction is based on an expectation, as for BLLiM, as well as the other methods we compare BLLiM with.

In this simulation setting, performances of each method are assessed 100 times: 100 independent learning datasets are generated to train the several compared models and the respective RMSE are computed through a testing dataset, generated independently from training data and consisting in $n=10~000$ observations. Empirical means and standard deviations of RMSE are therefore computed over those 100 repetitions. 


\subsubsection{Results} Results are presented in Table~\ref{tab:simus1}.
Unsurprisingly, BLLiM performs better than GLLiM as GLLiM ignores the dependence among covariates conditionally on the variable to predict. It illustrates that accounting for dependence in this inverse regression framework is crucial to achieve good prediction rates when covariates are correlated. BLLiM performs slightly better than GLLiM-$L_w$, which demonstrates that our block diagonal approach can achieve good prediction rates compared to GLLiM-$L_w$, while providing a more interpretable model from a biological point of view, as illustrated in Section~\ref{sec:realdata}. Finally, the number of observations does not seem to have a strong impact on the prediction results of the proposed method. It is interesting because it suggests that the block diagonal structure is correctly estimated, even when the number of observations is smaller than the true number of parameters to estimate, which is often the case on real data. The impact of sample size does not seem to have a strong impact on predictive performance of GLLiM, which is consistent as this model depends on a small number of parameters regarding to the two other compared methods. Moreover, the number of observations has a stronger impact on predictive performance of GLLiM-L$_w$ but this method remains competitive in prediction.

\begin{table}[ht]
\caption{Results of the simulation study: prediction error for 100 simulated datasets generated under locally affine regression functions for 2 different sample sizes. Three methods are compared, assuming that the large covariance matrix is diagonal  (GLLiM), block diagonal structured  (BLLiM) or under a factor decomposition  (GLLiM-L$_w$). Prediction accuracy is assessed by Root Mean Square Error  (mean and standard deviation).\label{tab:simus1}}
\begin{center}
\begin{tabular}{lcccc}
\hline
Number of observations & \multicolumn{2}{c}{ $n=4162$}&\multicolumn{2}{c}{ $n=416$}\\
 Method & Response 1 & Response 2& Response 1 & Response 2 \\
\hline
BLLiM & 0.078  (0.020) & 0.079  (0.015) & 0.105  (0.052) & 0.116  (0.098) \\
 GLLiM & 0.169  (0.062) & 0.152  (0.022) & 0.183  (0.059) & 0.168  (0.023) \\
GLLiM-$L_w$ & 0.072  (0.002) & 0.083  (0.002) & 0.136  (0.036) & 0.147  (0.038) \\
\hline
\end{tabular}
\end{center}
\end{table}

\subsection{Prediction of high-dimensional simulated manifolds\label{sec:simus2}}

In order to compare our procedure with other nonlinear regression methods of the literature, we consider some nonlinear link functions, relying on latent variables, with several covariance structures. It is built on the simulation plan of \cite{DFH14}.

\subsubsection{Simulation of high-dimensional manifolds}
\label{sec:simus2:framework}
We generate several datasets with covariates $\xv$ and a response $\yv= (t,\wv)$ which is partially observed, where $t$ is the observed part and $\wv$ is the latent part of the response. The covariates are related to the response through the following inverse model:
$$\xv = \fb (\yv) + \varepsilon = \fb (t,\wv) + \varepsilon$$
 where the different nonlinear functions $\fb$ considered in this paper are described hereafter and $\varepsilon$ is a Gaussian noise with several dependence structures detailed hereafter. These synthetic data are generated according to an inverse approach but the goal is to retrieve the observed part of the response $t$ from the covariates $\xv$. 

The phenotype response $\yv$ is multidimensional and partially observed: the quantitative trait $t$ is observed but $w_1$ and $w_2$ are two unobserved quantitative traits. Genes $\xv$ are simulated from $t$, $w_1$ and $w_2$, but we only seek to predict the observed quantitative trait $t$. However, the link between $t$ and $\xv$ also depends on the unobserved quantitative traits $w_1$ and $w_2$, which increases the complexity of the prediction problem on purpose.

We consider three different forms of inverse regression functions denoted by $\fb= (f_1,\ldots,f_D)$, $\gb= (g_1,\ldots,g_D)$ and $\hb= (h_1,\ldots,h_D)$ such as each component is, for $d\in\{1,\ldots,D\}$,
\begin{align*}
f_d (t,w_1) &=\alpha_d \text{cos}  (\eta_d t/10+\phi_d) + \gamma_dw_1^3; \\
g_d (t,w_1) &=\alpha_d \text{cos}  (\eta_d t/10+ \beta_d w_1 + \phi_d); \\
h_d (t,w_1,w_2) &= \alpha_d \text{cos}  (\eta_d t/10+ \beta_d w_1 + \phi_d)+ \gamma_dw_2^3.
\end{align*}
The vector of covariates $\xv= (x_1,\ldots, x_D)$ has dimension $D=50$ and is generated dimension by dimension through one of each nonlinear function described above. 

The observed response $t$ is uniformly sampled in $[1,10]$ and the hidden responses $ (w_1,w_2)$ are uniformly sampled in $[-1,1]$. For each of the three functions, 100 training and 100 testing datasets are generated with $N=200$ observations. For each of the 100 runs, 
a set of parameters is uniformly sampled such as $\alpha_d \in [0,2 ]$, $\eta_d \in [0,4\pi ]$, $\phi_d \in [0,2\pi ]$, $\beta_d \in [0,\pi ]$ and $\gamma_d \in [0,2]$ and used to generate the $N=200$ profiles of covariates. This means that the difficulty of the problem differs at each run. This simulation plan allows to cover a large range of situations. Moreover, adding unobserved responses $\wv$ is interesting to study the sensibility of BLLiM to the presence of partially observed response. 

\begin{figure}[ht!]
\begin{center}
\caption{Image of correlations matrices $ (\Sigmab_k)_{1\leq k \leq K}$ used in simulations of Section~\ref{sec:simus2}
 \label{fig:matrices}}
\includegraphics[width=4.1cm]{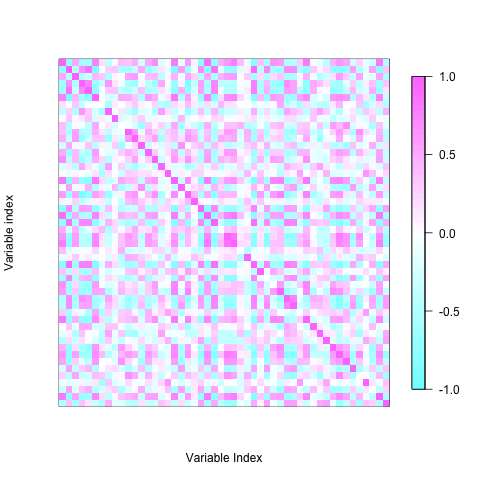}
\includegraphics[width=4.1cm]{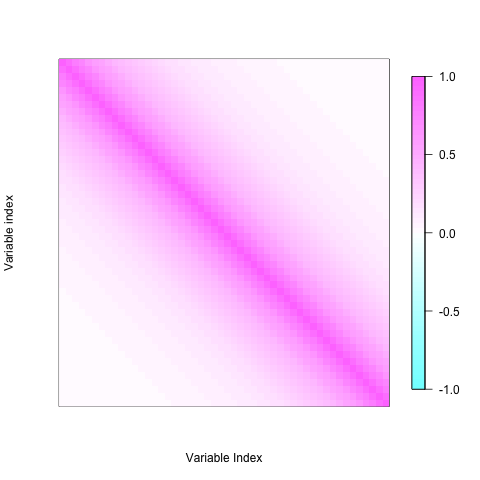}
 \includegraphics[width=4.1cm]{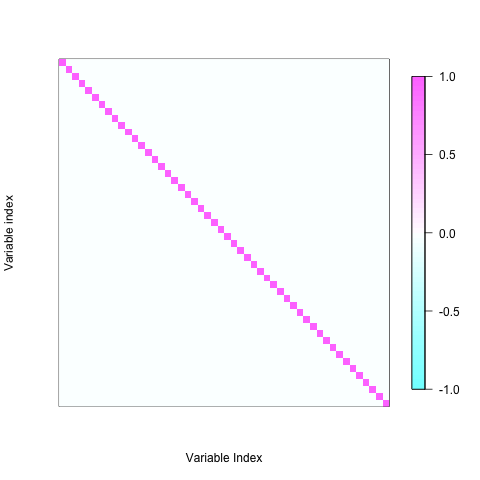}
\includegraphics[width=4.1cm]{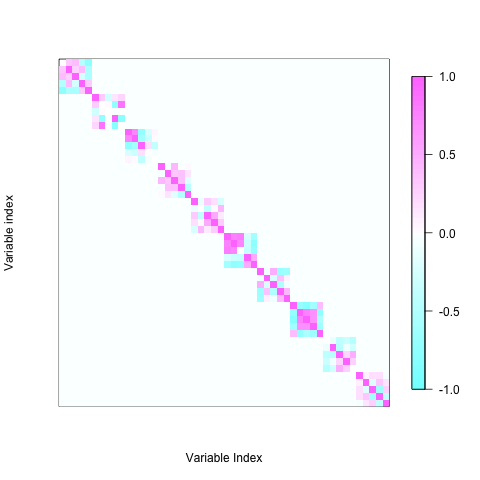}
\end{center}
\end{figure}

We consider several dependence structures for $ (\Sigmab_k)_{1\leq k \leq K}$ described hereafter and displayed on Figure~\ref{fig:matrices}:
\begin{itemize}
\item Factor structure: each $\Sigmab_k$ is decomposed as $\Phib_k + \Cb_k\Cb^T_k$ where $\Phib_k$ is a $D$-diagonal matrix and $\Cb_k$ is a $D \times q$ matrix, with $q=5$. For each run of the simulations, in this sparse and low rank decomposition, $\Phib_k$ and $\Cb_k$ are randomly generated such as $\Tr
 (\Cb\Cb^T)/\Tr (\Sigmab)=0.9$, which mimics general strong dependence patterns  (see for example \cite{friguet09,leekStorey07,leekStorey08,perthame15} for more details). The matrix is normalized to be a correlation matrix. This design is favorable to GLLiM-$L_w$ as latent factors are underlying in the structure of $\Sigmab_k$. 
\item Toeplitz structure: each $\Sigmab_k$ is a Toeplitz matrix whose term indexed by $(i,j)$ is $ (\Sigmab_k)_{i,j}=0.9^{\vert i-j\vert}$. This design is also a strong temporal dependence pattern of an autoregressive process. This design is not favorable 
for both BLLiM and GLLiM-L$_w$.
\item Independence structure: each $\Sigmab_k$ is set to the identity.
\item Blocks structure: $\Sigmab_k$ contains 10 blocks of 5 variables. Factor decomposition  ($\Phib_k+\Cb\Cb^T_k$) holds on in each block with a strong dependence structure. Each block is independent from the others and independently generated among clusters. 
\end{itemize}

The prediction accuracy of those methods is assessed by the RMSE described in \eqref{def::MSE}.

\subsubsection{Compared methods} \label{ComparedMethod} In this simulation study, we propose to compare our method to model-based and non model-based prediction methods of the literature. The following methods are compared: 
\begin{itemize}
\item Model-based approaches
\begin{itemize}
\item BLLiM: the proposed method described in Section \ref{sec:sel}, where both $K$ and the block diagonal structures of $ (\Sigmab_k)_{1\leq k \leq K}$ are selected by the slope heuristic;
\item GLLiM: standard version of GLLiM with heteroscedastic noise  ($ (\Sigmab_k)_{1\leq k \leq K}$ are diagonal). The number of affine components $K$ is selected by minimizing BIC criterion, as proposed in \cite{DFH14}; 
\item GLLiM-L$_w$: hybrid version of GLLiM, in which $L_w$ latent factors can be added to complete the response when it is partially observed, both $ (K,L_w)$ are selected by minimizing BIC criterion, as proposed in \cite{DFH14};
\item MARS: multiple adaptive regression splines proposed by \cite{F91} and implemented in the \texttt{mda R} package;
\item SIR: sliced inverse regression, introduced by \cite{L91}, followed by polynomial regression with polynomial function of order 3. Results are presented for 2 directions as it is the number of slices that provides the best results. SIR directions are extracted using the \texttt{dr R} package;
\item nonlinear PLS proposed by \cite{WKS89} and implemented in the \texttt{ppls R} package. Parameters are tuned by cross-validation  (CV) as proposed in the package;
\item K-plane proposed by \cite{MS15}. This method is based on an approach similar to GLLiM and estimation is based on EM-algorithm. We present the results for $K=2$ as this number provides the best results for this method. A Matlab code implementing the method is available on 
 \href{https://services.math.duke.edu/~yiwang/KMapRT.html}{https://services.math.duke.edu/$\sim$yiwang/KMapRT.htm}.
\end{itemize}
\item Non model-based approaches\\
Even if these competitive methods do not provide interpretable results from a biological point of view, they are known to achieve very good prediction results so it could be interesting to compare our model-based approach in terms of prediction accuracy. Therefore, our procedure is compared to the following methods: 
\begin{itemize}
\item Random forest \cite{B01} using \texttt{randomForest R} package;
\item SVM: support vector machine \cite{V98} using \texttt{e1071 R} package, with a Gaussian kernel;
\item RVM: relevance vector machine proposed by \cite{T01} and implemented in the \texttt{kernlab R} package, with a linear kernel.
\end{itemize}
\end{itemize}

\subsubsection{Results}
Table \ref{TableVariete2} presents the results of the procedures described in Section \ref{ComparedMethod} for the three nonlinear regression functions and for the four several covariance structures described in Section \ref{sec:simus2:framework}.

\begin{table}[ht!]
\footnotesize
\begin{center}
\caption{Results of simulations study: prediction errors computed on datasets simulated under nonlinear manifolds and described in Section \ref{sec:simus2}. Latent variables are hidden and one response is predicted. Several model-based and non model-based regression methods are compared. Prediction accuracy is assessed by computing Root Mean Square Error on independent testing datasets. Each method is assessed 100 times. The standard deviations are also computed. Results better than our procedure are highlighted in bold.
}
\begin{tabular}{lcccccc}
\\
\hline 
&\multicolumn{3}{c}{\textbf{Factor structure}}&\multicolumn{3}{c}{\textbf{Toeplitz structure}} \\
& f & g & h& f & g & h\\
\hline
\multicolumn{3}{l}{\textit{Model-based methods}}&&\\
BLLiM
& 0.965  (0.333) & 1.749  (0.359) & 1.927  (0.353) & 1.097  (0.402) & 1.725  (0.487) & 2.025  (0.498) \\
GLLiM & 1.205  (0.321) & 2.201  (0.389) & 2.250  (0.371) & 1.598  (0.269) & 2.417  (0.380) & 2.495  (0.401) \\ 
GLLiM-L$_w$ & \textbf{0.841  (0.314)} & 1.749  (0.384) & \textbf{1.838  (0.342)} & \textbf{1.049  (0.208)} & {1.846  (0.346)} & \textbf{1.920  (0.358)} \\
MARS & 1.039  (0.209) & 1.873  (0.416) & \textbf{1.790  (0.377)} & \textbf{0.796  (0.154)} &\textbf{1.490  (0.375)} & \textbf{1.580  (0.364)} \\
SIR
& 1.497  (0.584) & 1.771  (0.537) & \textbf{1.712  (0.537)} & 1.282  (0.569) &\textbf{1.582  (0.615) }&\textbf{1.594  (0.581)} \\
nonlinear PLS & \textbf{0.763  (0.136)} & \textbf{1.388  (0.286)} & \textbf{1.350  (0.289)} & \textbf{0.580  (0.102)} & \textbf{1.116  (0.253)} & \textbf{1.182  (0.266)} \\
Kplane & 1.079  (0.478) & \textbf{1.446  (0.703)} & \textbf{1.421  (0.636) }&\textbf{ 0.917  (0.388)} & \textbf{1.272  (0.585)} & \textbf{1.308  (0.664)}\\ 
\multicolumn{3}{l}{\textit{Non model-based methods}}&&\\
randomForest & 1.124  (0.196) & \textbf{1.595  (0.263)} & \textbf{1.644  (0.260)} & \textbf{1.081  (0.155)} &\textbf{1.541  (0.246) }&\textbf{1.629  (0.250)} \\
SVM 
& 1.006  (0.135) & \textbf{1.404  (0.200)} & \textbf{1.449  (0.186)} & \textbf{0.925  (0.116)} & \textbf{1.334  (0.192)} &\textbf{1.392  (0.199)} \\
RVM 
& 1.366  (0.229) & 2.394  (0.539) & 2.535  (0.583) & \textbf{0.997  (0.174)} &2.004  (0.552) &\textbf{1.949  (0.509)} \\
\hline
&\multicolumn{3}{c}{\textbf{Independence structure}} &\multicolumn{3}{c}{\textbf{Blocks structure}}\\
& f & g & h& f & g & h\\
\hline
\multicolumn{3}{l}{\textit{Model-based methods}}&&\\
BLLiM
& 0.743  (0.305) & 1.301  (0.286) & 1.556  (0.343) & 0.658  (0.273) & 1.221  (0.308) & 1.564  (0.342) \\
GLLiM & \textbf{0.710  (0.272)} & 1.553  (0.248) & 1.841  (0.324) & 0.777  (0.254) & 1.566  (0.301) & 1.952  (0.346) \\
GLLiM-L$_w$ & \textbf{0.640  (0.170)} & 1.463  (0.287) & 1.603  (0.297) & 0.710  (0.176) & 1.451  (0.326) & 1.704  (0.304) \\
MARS & 1.620  (0.279) & 2.367  (0.401) & 2.439  (0.434) & 1.360  (0.256) & 2.115  (0.404) & 2.295  (0.406) \\
SIR
 & 1.888  (0.571) & 2.339  (0.417) & 2.342  (0.408) & 1.731  (0.642) & 2.044  (0.504) & 2.113  (0.442)\\
nonlinear PLS & 1.205  (0.172) & 1.878  (0.242) & 1.898  (0.257) & 1.061  (0.183) & 1.732  (0.331) & 1.814  (0.293) \\
Kplane & 1.412  (0.579) & 1.714  (0.618) & 1.739  (0.659) & 1.262  (0.552) & 1.601  (0.693) & 1.624  (0.668) \\ 
\multicolumn{3}{l}{\textit{Non model-based methods}}&&&&\\
randomForest & 1.205  (0.219) & 1.704  (0.221) & 1.798  (0.245) & 1.234  (0.224) & 1.681  (0.290) & 1.833  (0.237) \\
SVM
& 1.154  (0.158) & 1.612  (0.159) & 1.684  (0.188) & 1.122  (0.164) & 1.545  (0.210) & 1.658  (0.198) \\
RVM
 & 2.413  (0.378) & 3.694  (0.592) & 3.668  (0.652) & 2.011  (0.397) & 3.229  (0.708) & 3.252  (0.612)\\
\hline
\label{TableVariete2}
\end{tabular}
\end{center}
\end{table}

Note that the regression functions $\gb$ and $\hb$ are harder to estimate than $\fb$, due to the hidden variables playing a nonlinear role. This design of partially observed response is difficult to catch. In contrast, $\fb$ is a linear function in hidden variables, and the design of partially observed response is easier to catch for every method.

As expected, BLLiM outperforms other methods on data simulated with a block diagonal residual covariance matrix. Indeed, BLLiM is able to identify the underlying block diagonal structure of correlation matrices in order to improve prediction rates. 

It is interesting to note that BLLiM performs well on the independent structure, which suggests that BLLiM does not impose non-zero correlation structure and detect small blocks, when predictors are independent. BLLiM outperforms other methods for the regression functions $\gb$ and $\hb$.

On data simulated with the factor structure, BLLiM is competitive with other methods. Nonlinear PLS and SVM perform better, but BLLiM achieves good prediction accuracy. For example, it performs as well as GLLiM-L$_w$, which is designed to catch such correlation matrices. It is interesting to notice that the block structure succeed to adapt itself to model the factor structure.

Remark that the Toeplitz design is harder to catch by BLLiM procedure. Performances are outperformed by nonlinear PLS, Kplane, and SVM.

Finally, remark that MARS and the nonlinear PLS achieve similar prediction results than BLLiM, or even better in some cases which suggests that it is challenging to improve their prediction performance.
However, the main advantage of BLLiM is that it provides an interpretable model, which will be highlighted in Section \ref{sec:realdata}.

\section{Application on the prediction of olfactory behaviour from transcriptomic data\label{sec:realdata}}
\label{RealData}
 In this section, the performances and the interpretability of our model are illustrated on a study that aims at understanding the biological variation of olfactory behaviour in drosophila, an animal model widely used in genetics \cite{Roberts2006}. To this end, we use the ressources available from the Drosophila Melanogaster Reference Panel Project  (DGRP) \cite{Mackay2012}, a community resource developed to analyze population genomics and quantitative traits. 

Understanding natural variation in olfactory behaviour can help to understand survival and reproduction. \cite{Swarup2013} have measured a response index to exposure of benzaldehyde for 163 inbred drosophila lines from the DGRP for males and females. A response index of 1 corresponds to a maximal aversive response to the benzaldehyde source whereas a response index of 0 means that all flies remain close to source. To understand natural variation in olfactory behaviour, \cite{Swarup2013} have performed three complementary genome-wide association studies. In another article, \cite{Huang2015} have described how the variation in RNA level is a link between the variation at the DNA level and the phenotype. We have downloaded the response index to benzaldehyde and the transcriptomic data for each inbred drosophila lines  (male and female) from the following link \href{http://dgrp2.gnets.ncsu.edu/data.html}{http://dgrp2.gnets.ncsu.edu/data.html}. In \cite{Huang2015}, measurement of genes expression have been performed for 18140 genes and 205 inbred drosophila lines, for males and females. Two replicates are available for the transcriptomic dataset, only one replicate was available for the response index to benzaldehyde, for each inbred line. We took the mean value of the two replicates for the transcriptomic dataset. To preselect genes, we first retained the top $M=4000$ highest variable genes. For each remaining genes, we computed a model independent measure of variable importance using a non parametric method described in \cite{Kuhn2008}, Section 9: a LOESS smoother is fit between the response index and each gene expression. For an univariate response, a coefficient of determination $R^2_d$ is computed as follows:
\begin{eqnarray}
R^2_d &= 1 - \frac{\sum_{i=1}^n  (y_i - \hat y_{i,d})^2}{\sum_{i=1}^n  (y_i - \bar{y})^2} \nonumber
\end{eqnarray}
where $y_i$ is the observed response of subject $i$, $\hat y_{i,d}$ is the corresponding prediction by a LOESS regression with only predictor $d$ and $\bar{y}_i$ is the mean of responses $ (y_1,\ldots,y_n)$. The collection of coefficients of determination $ (R_1^2,\ldots,R_M^2)$ is used as a relative measure of variable importance of each gene to predict the olfactory behaviour. We kept the top 50 genes with the highest measure of importance. The final dataset consists in one quantitative trait variable  ($L=1$) and transcriptomic data for 50 genes  ($D=50$), collected on the same inbred drosophila  ($n=326$). The dataset also contains one variable: the sex of each subject, male or female. This variable is not included to perform prediction and is used for downstream biological interpretation of the analysis. We performed prediction using  BLLiM procedure on this dataset. The  range of clusters $\mathcal{K} = \{2,\ldots,4\}$ has been tested. Increasing the number of clusters leads to empty clusters. To select the number of clusters, we have used BIC instead of the slope heuristic because there are not enough points to calibrate the coefficient in the slope heuristic.

In the following subsection, our model is compared by cross-validation with other prediction strategies to demonstrate the good predictive performance of our method. In the final section, the clustering of individuals performed to catch the nonlinear relationship between the quantitative trait to predict and the transcriptomic data is analyzed. Subsequently, we explore and compare the gene regulatory networks inferred in each cluster of individuals.

\subsection{Assessment of prediction accuracy}

In this section, our prediction strategy is compared to other prediction strategies by a 10-fold cross validation procedure. Predictive performances of the methods are measured by the RMSE, defined in \eqref{def::MSE}. More precisely, we compute the RMSE for each run of the 10-fold cross validation procedure, and we repeat it 50 times to consider statistics of this criterion. Each method is therefore assessed 500 times. The results are presented in Table~\ref{tab:drosoCV}. Our procedure achieves challenging prediction performances, similar to nonlinear PLS and non model-based methods. Non model-based methods are known to perform well in difficult setting and real data but the downstream model interpretation is not clear. In nonlinear PLS, the nonlinear relationship between latent variables is not as easy to interpret  as latent factors do not represent any biological entities. Moreover, we outperform the prediction error of MARS, SIR and K-plane. 
 For example, K-plane has very good performances on simulated dataset, as described in Table \ref{TableVariete2}, but perform poorly on this real dataset. Note that, for the K-plane prediction, the authors do not propose a criterion to choose the number of components. We choose the number of components that minimizes the prediction error on testing data, which is rather optimistic as testing data should not be used to tune hyper parameters.

\begin{table}[ht!]
\begin{center}
\caption{Errors computed by 10-fold cross validation for the prediction of response index to benzaldehyde \cite{Swarup2013} from transcriptomic data \cite{Huang2015} from data collected in the Drosophila Melanogaster Reference Panel Project \label{tab:drosoCV}. Mean and standard deviation are computed over 50 repetitions of the cross validation process. Results better than our procedure are highlighted in bold.}
\begin{tabular}{lccc}
\hline
Method & RMSE  (sd) \\
\hline
\textit{Model-based methods} & \\
BLLiM & 0.0798  (0.002) \\
GLLiM & 0.0970  (0.003) \\
GLLiM-$L_w$ & 0.0850  (0.001) \\
MARS & 0.0924  (0.003) \\
SIR & 0.0836  (0.001) \\
nonlinear PLS & \textbf{0.0748  (0.001)} \\
K-plane & 0.1039  (0.004) \\
\textit{Non model-based methods} & & & \\
Random Forest & \textbf{0.0764  (4e-4)} \\
SVM - Gaussian kernel & \textbf{0.0764  (5e-4)} \\
RVM - Linear kernel & \textbf{0.0742  (0.001)} \\
\hline
\end{tabular}
\end{center}
\end{table}

\subsection{Interpretation of BLLiM model}

To approximate the nonlinear function which relates the phenotypic variable to the expression data, BLLiM separates drosophila into 3 clusters of sizes 
 163, 106 and 57. In fact, cluster 1 corresponds to male drosophila  (in red in the plots), whereas clusters 2 and 3 correspond to female drosophila  (in green and blue in the plots). The correspondence of these inferred clusters and the external information  (sex) demonstrates the ability of our model to take into account hidden
observations stratification: even though the sex information is not included into the model, the proposed method is able to retrieve this information from the data.

\begin{figure}[ht!]
\caption{Response index to benzaldehyde for each cluster of drosophila  (Left). Each color represents a cluster of drosophila  (red for cluster 1, green for cluster 2 and blue for cluster 3). We remark that there is no difference in the mean level between the clusters. The difference relies on the link between genes and response index to benzaldehyde, as illustrated by the representation of the absolute value of the 50 coefficients of the linear regression to predict the response index to benzaldehyde from the gene expression data for each cluster of drosophila  (Right). }
\begin{center}
\begin{minipage}{7.5cm}
\includegraphics[height=5.5cm, trim = 0cm 0cm 0.5cm 1.5cm, clip]{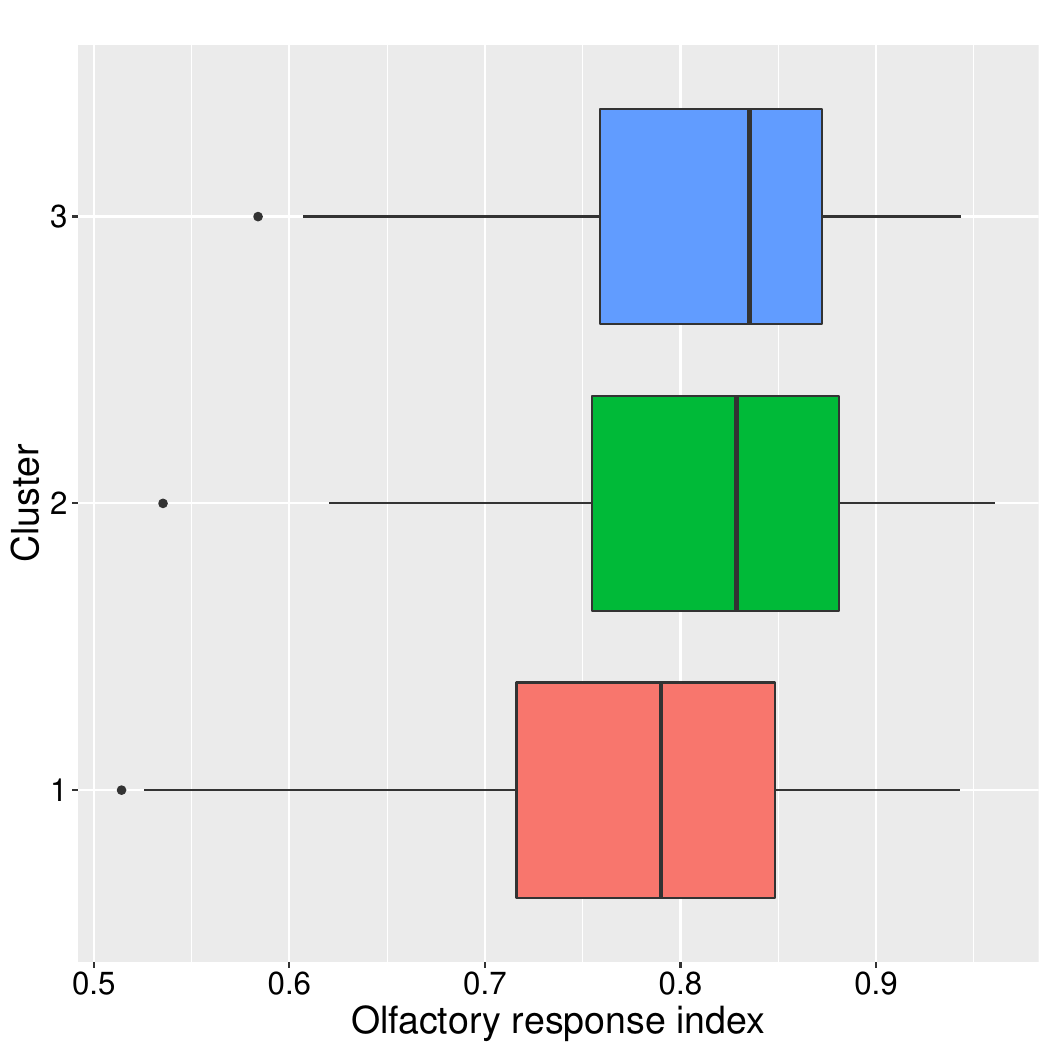}
\end{minipage}
\begin{minipage}{6cm}
\includegraphics[height=6cm, trim = 0cm 0cm 0cm 0cm, clip]{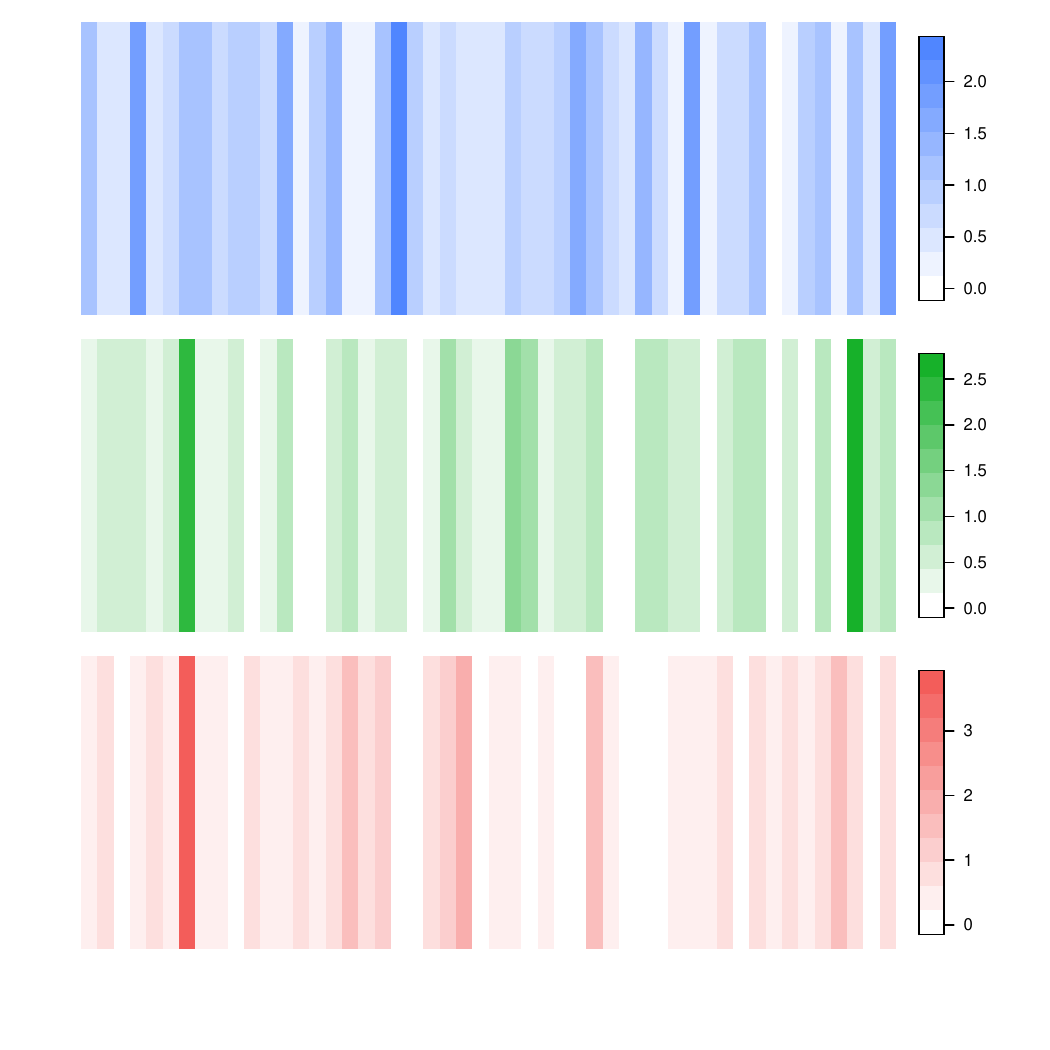}
\end{minipage}
\end{center}
\label{boxplotCluster}
\end{figure}

The response index to benzaldehyde in each cluster of drosophila is represented in Figure \ref{boxplotCluster}  (Left). We notice no significant differences between the response index among the clusters. Indeed, the clustering is based on the relation between response and covariates, and not on the value of the response. In contrast, the values 
taken by coefficients in the linear regression are highly different between clusters, as illustrated in Figure \ref{boxplotCluster}  (Right). These differences demonstrate the necessity of taking nonlinearity into account. The genes contributing the most to the prediction of response index to benzaldehyde differ between clusters. 

\begin{figure}[ht!]
\begin{center}
\caption{Modules of genes  (nodes) detected for each cluster of individuals. The name of the genes is written on each node. The edges correspond to interactions between genes. The color of the edges indicates the cluster of individuals for which the interaction is observed: blue, green and red edges correspond to cluster 1, 2 and 3 respectively. The color of the node is yellow if the gene is shared among modules detected both for cluster 1, 2 and 3. The software Cytoscape has been used to plot the graphs \cite{Rowan2005}.}
\label{petitCluster}
\includegraphics[width=4.8cm, trim = 0.1cm 1.5cm 5cm 5.5cm, clip]{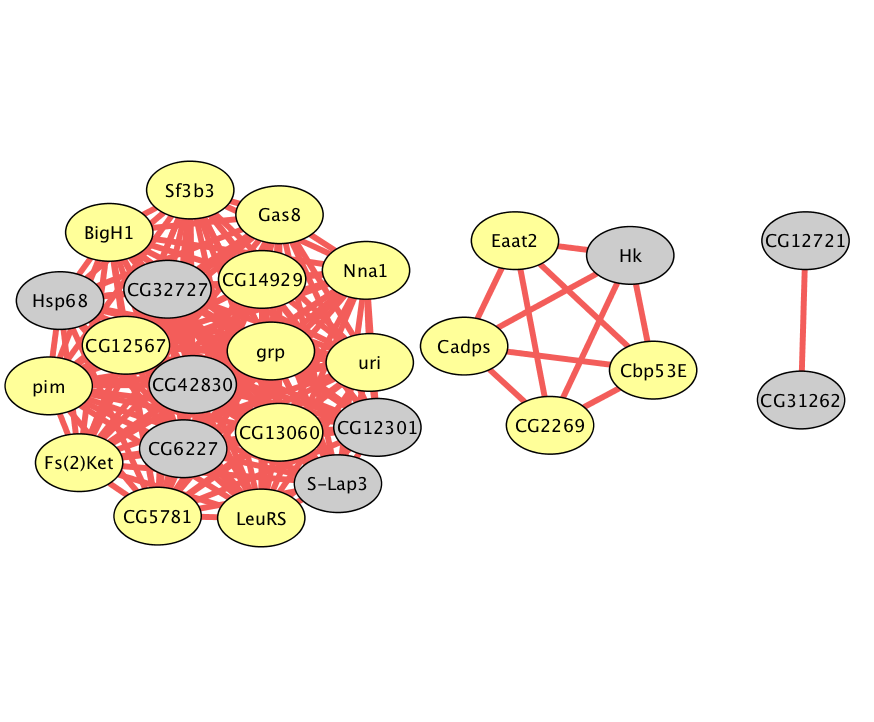}
\includegraphics[width=4.8cm, trim = 0cm 0.5cm 10cm 0cm, clip]{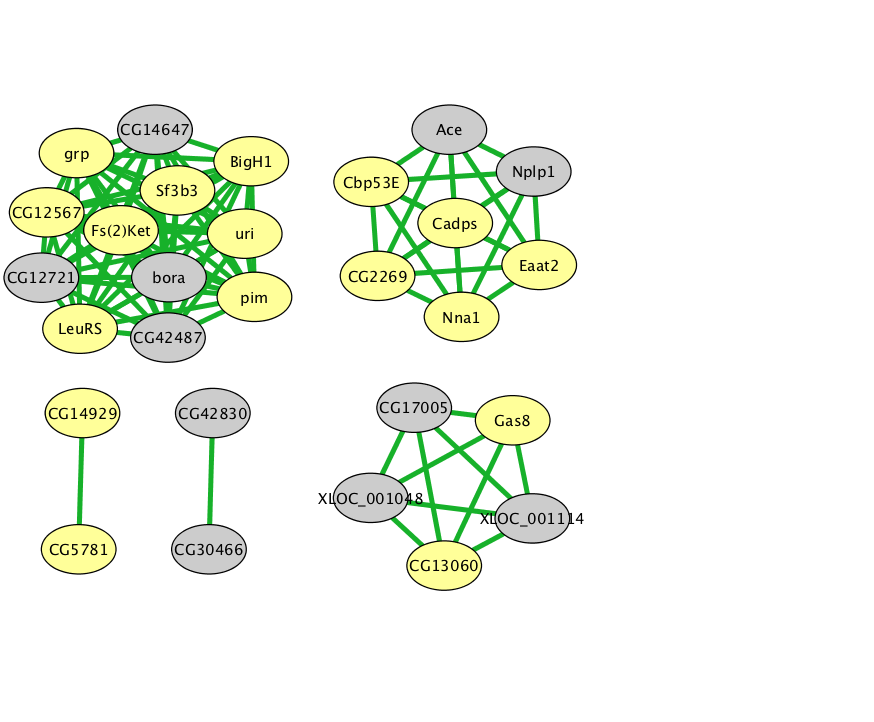}
\includegraphics[width=6.8cm, trim = 0cm 0.5cm 0cm 2.3cm, clip]{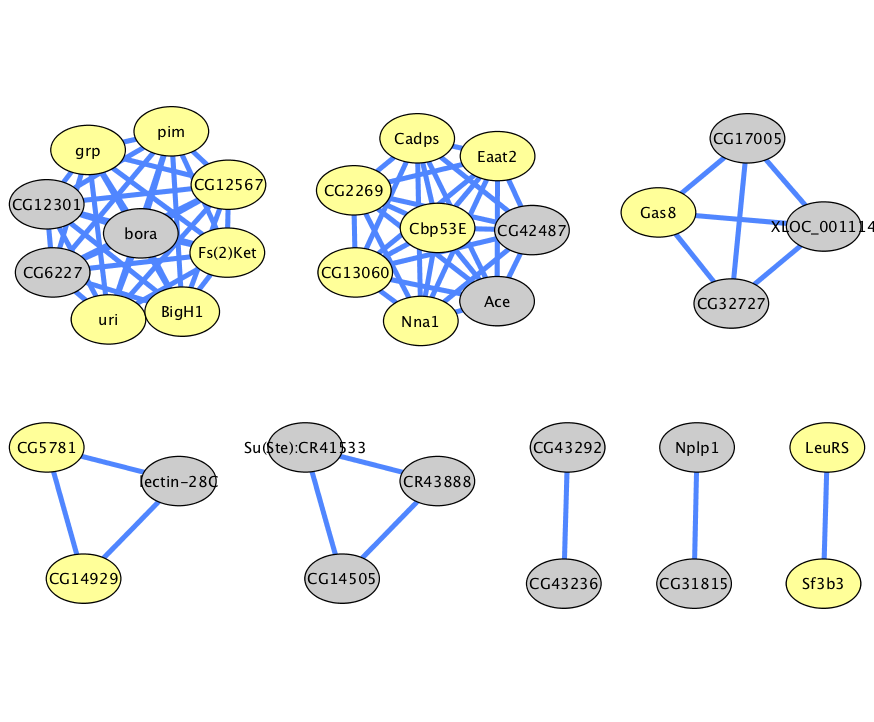}
\end{center}
\end{figure}

In a similar way, the modules of the inferred gene regulatory network are not similar among clusters. 
For cluster 1  (males), 3 modules of size 19, 5 and 2 are detected. For cluster 2, 5 modules of size 12, 7, 5, 2 and 2 are detected. For cluster 3, 8 modules of size 9, 8, 4, 3, 3, 2, 2, and 2 are detected. The sizes of the modules are smaller than the $50$ genes we consider. The corresponding modules are represented in Figure \ref{petitCluster}. 
We obtain a sparse representation of the network underlying the phenotype transcriptome relationship for each cluster of individuals. Remark that isolated genes are not displayed in Figure~\ref{petitCluster}.

\begin{figure}[ht!]
\begin{center}
\caption{A merged version of the networks inferred for each cluster of individuals is presented  (Right). The name of the genes is written on each node. The edges correspond to interactions between genes. The color of the edges indicates the cluster of individuals for which the interaction is observed. Blue, green and red edges correspond to cluster 1, 2 and 3 respectively. The genes common to all clusters are colored in yellow. The genes isolated for every cluster of individual are colored in red. The software Cytoscape has been used to plot the graph \cite{Rowan2005}.}
\includegraphics[width=12.5cm, trim = 0cm 5.5cm 0cm 6cm, clip]{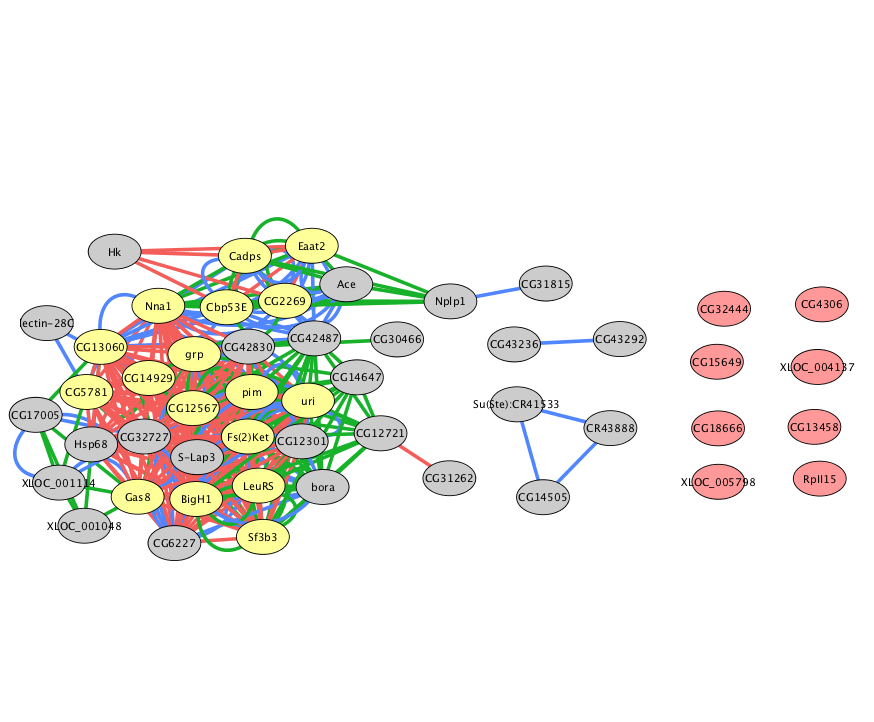}
\label{VennBigGraph}
\end{center}
\end{figure}

 A merged version of the 3 different networks presented in Figure \ref{petitCluster} is presented in Figure \ref{VennBigGraph}. 
 A core set of 17 genes is detected in a network in every cluster. These genes are highlighted in yellow in Figures \ref{petitCluster} and \ref{VennBigGraph}.  Few genes belong to exactly two clusters: 2 genes between clusters 1 (red) and 2 (green), 6 genes between clusters 2 (green) and 3 (blue), and 3 genes between clusters 1 (red) and 3 (blue), and some are detected only in one cluster.

Among the list of 17 genes in modules for each cluster of individuals, we notice that genes \textit{Sf3b3} and \textit{LeuRS} are always interconnected, for cluster 1  (red), cluster 2  (green) and cluster 3  (blue). Genes \textit{Sf3b3} and \textit{LeuRS} are known to be implicated in neurogenesis \cite{Neumuller2011}, and olfactory behaviour is known to be linked with neurogenesis cellular processes. This observation is consistent with the GWA studies performed in \cite{Swarup2013}. Similarly, genes \textit{Cadps} and \textit{Cpb53E} are always interconnected, in cluster 1  (red), cluster 2  (green) and cluster 3  (blue). This results is not surprising either, given that agressive behaviour has an olfactory component and that \textit{Cadps} and \textit{Cpb53E} are known to be associated with neurotransmitter secretion \cite{Lloyd2000} and negative regulation of response to wounding \cite{Wishart2012}, respectively. Gene \textit{Hk}, found in the inferred network for cluster 1  (males), is known to be implicated in flight behaviour \cite{Homyk1977}. Conversely, gene \textit{Nplp1}, is known to be implicated in neuropeptide signaling pathway \cite{Baggerman2002}, is only found in modules for cluster 3  (cluster of 57 females). Genes \textit{pim}  (in modules for all clusters) and \textit{Hsp68}  (in modules for cluster 1 only) are also known to be implicated in neurogenesis \cite{Neumuller2011}. 
Among the 17 genes in modules for each cluster of individuals, we also notice gene \textit{Gas8}, known to be implicated in process involved in the controlled movement of a flagellated sperm cell, gene \textit{Fs (2)Ket}, known to be linked with the construction of a chorion-containing eggshell \cite{Schupbach1991}, and gene \textit{grp}, known to be implicated in female meiosis chromosome segregation \cite{Dobie2001}. These genes are linked with reproduction, which is known to have an olfactory component. Several genes identified in modules for all clusters, such as \textit{CG13060} and \textit{CG2269} have not been identified before as linked to the variation in olfactory behaviour in adult drosophila and can be investigated in further studies. These results, based on the phenotype-transcriptome relationship study, may complete classical GWA studies and help to improve the understanding of the natural variation of olfactory behaviour.

The code to reproduce the complete real data analysis has been provided as supplementary materials.

\section{Discussion\label{sec:conclu}}

In this paper, an interpretable statistical framework is proposed to perform non linear prediction from transcriptomic data. Our prediction method performs well on simulated and real data and is competitive with other prediction models, including machine learning techniques. Unlike other methods, the proposed model is fully parametric and interpretable: the nonlinear relationship between the phenotypical variables to predict and the transcriptome data is modeled by locally linear regressions, clusters of individuals are encoded by hidden variables associated with a mixture of linear regressions model and the relationship between genes is encoded by a block structure in the conditional covariance matrices of the model.

In this paper, our method has been used on a real dataset with 50 covariates. Considering the number of parameters with respect to the number of observations, the limits of BLLiM would be reached if the number of variables $D$ grows to thousands while the number of observations $N$ remains around hundreds, which is an usual design in genomics studies. A standard idea in this context of unbalanced design is to regularize the estimation of regression coefficients in order to reduce dimension. Introducing a sparse structure in the regression coefficients of the inverse regression, namely the collection of matrices $ (\Ab_k)_{1 \leq k \leq K}$ would be particularly relevant in the BLLiM framework. It is interesting to notice that the estimated matrices $ (\widehat \Ab_k)_{1 \leq k \leq K}$ displayed on Figure~\ref{boxplotCluster}  (Right) exhibit a sparse structure which would be a valuable information to account for in the model. 

Besides, the proposed method is based on a mixture of linear regressions, which allows to estimate gene regulatory networks within each cluster of the mixture. Both on biological and statistical point of view, it could be valuable to include in the model a regulatory network common to all clusters, and cluster-specific networks, in the manner of the joint graphical lasso \cite{Danaher2014}. Some genes would therefore be shared among clusters of individuals : some modules would be shared across different clusters of individuals  (functional modules), some modules would be specific for each cluster  (disease modules).

In this paper, the BLiMM method is applied on transcriptomic data to study the olfactory behaviour of drosophila. As going from the genome level to the transcriptome level helps to gain deeper insights into complex biological diseases such as cancers \cite{Rhodes2005}, an application of our model on cancer data would be of particular interest in the context of network medicine, as described by \cite{Barabasi2011}: modules detected for each set of individuals may correspond to functional modules or disease modules. The performance of the method is demonstrated on transcriptomic data but other "omics" data can be used, such metabolomic data or proteomic data. Our model would also be useful in the context of trans"omics" data \cite{Yugi2016}, combining data from multiple "omics" layers to understand deeper biological processes. 

Finally, our prediction framework can be used to perform prediction on any dataset with heterogeneous observations and hidden graph-structured interactions between covariates. This framework is of interest beyond molecular biology, and can be useful in other fields (neurology, ecology or sociology). 

\section*{Acknowledgements}
 Emilie Devijver is supported by the Interuniversity Attraction Poles Programme  (IAP-network P7/06), Belgian Science Policy Office, whose funding is gratefully acknowledged. We thank Jordi Estell\' e  (INRA, Jouy-en-Josas, France) for suggesting the Drosophila Melanogaster Genetic Reference Panel as a ressource for real data analysis.

\bibliographystyle{apalike}
\bibliography{biblio}

\begin{thebibliography}{}

\bibitem[Arlot and Massart, 2009]{Arlot2010}
Arlot, S. and Massart, P. (2009).
\newblock Data-driven calibration of penalties for least-squares regression.
\newblock {\em Journal of Machine Learning Research}, 10:245--279.

\bibitem[Arumugam and Scott, 2004]{AS04}
Arumugam, M. and Scott, S. (2004).
\newblock Emprr: A high-dimensional em-based piecewise regression algorithm.
\newblock In {\em Proceedings of International Conference on Machine Learning
  and Applications}, pages 264--271.

\bibitem[Baggerman et~al., 2002]{Baggerman2002}
Baggerman, G., Cerstiaens, A., {De Loof}, A., and Schoofs, L. (2002).
\newblock {Peptidomics of the larval Drosophila melanogaster central nervous
  system.}
\newblock {\em Journal of Biological Chemistry}, 277(43):40368--40374.

\bibitem[Barab{\'{a}}si and Albert, 1999]{Barabasi1999}
Barab{\'{a}}si, A.~L. and Albert, R. (1999).
\newblock {Emergence of scaling in random network}.
\newblock {\em Science}, 286:509--512.

\bibitem[Barab{\'{a}}si et~al., 2011]{Barabasi2011}
Barab{\'{a}}si, A.~L., Gulbahce, N., and Loscalzo, J. (2011).
\newblock {Network medicine: a network-based approach to human disease}.
\newblock {\em Nature Reviews Geneticss}, 12(1):56--68.

\bibitem[Barab{\'{a}}si and Oltvai, 2004]{Barabasi2004}
Barab{\'{a}}si, A.~L. and Oltvai, Z.~N. (2004).
\newblock {Network Biology: Understanding the Cell's Functional Organization}.
\newblock {\em Nature Reviews Genetics}, 5:101--113.

\bibitem[Baraud et~al., 2009]{Baraud}
Baraud, Y., Giraud, C., and Huet, S. (2009).
\newblock Gaussian model selection with an unknown variance.
\newblock {\em The Annals of Statistics}, 37(2):630--672.

\bibitem[Baudry et~al., 2012]{Baudry}
Baudry, J.-P., Maugis, C., and Michel, B. (2012).
\newblock Slope heuristics: overview and implementation.
\newblock {\em Statistics and Computing}, 22(2):455--470.

\bibitem[Birg\'e and Massart, 2001]{BirgeMassart}
Birg\'e, L. and Massart, P. (2001).
\newblock Gaussian model selection.
\newblock {\em Journal of the European Mathematical Society}, 3(3):203--268.

\bibitem[Bouveyron et~al., 2015]{Bouveyron}
Bouveyron, C., C\^ome, E., and Jacques, J. (2015).
\newblock The discriminative functional mixture model for a comparative
  analysis of bike sharing systems.
\newblock {\em The Annals of Applied Statistics}, 9(4):1726--1760.

\bibitem[Breiman, 2001]{B01}
Breiman, L. (2001).
\newblock Random forests.
\newblock {\em Machine Learning}, 45(1):5--32.

\bibitem[Butte et~al., 2000]{Butte2000}
Butte, A.~J., Tamayo, P., Slonim, D., Golub, T.~R., and Kohane, I.~S. (2000).
\newblock {Discovering functional relationships between RNA expression and
  chemotherapeutic susceptibility using relevance networks.}
\newblock {\em Proceedings of the National Academy of Sciences},
  97(22):12182--12186.

\bibitem[Cand\`es et~al., 2009]{robustPCA}
Cand\`es, E.~J., Li, X., Ma, Y., and Wright, J. (2009).
\newblock Robust principal component analysis?
\newblock {\em Journal of ACM}, 58(1):1--37.

\bibitem[Chandrasekaran et~al., 2011]{lowrank11}
Chandrasekaran, V., Sanghavi, S., Parrilo, P., and Willsky, A. (2011).
\newblock Rank-sparsity incoherence for matrix decomposition.
\newblock {\em SIAM Journal on Optimization}, 21(2):572--596.

\bibitem[Christmas et~al., 2005]{Rowan2005}
Christmas, R., Avila-Campillo, I., Bolouri, H., Schwikowski, B., Anderson, M.,
  Kelley, R., Landys, N., Workman, C., Ideker, T., Cerami, E., Sheridan, R.,
  Bader, G.~D., and Sander, C. (2005).
\newblock {Cytoscape: a software environment for integrated models of
  biomolecular interaction networks}.
\newblock {\em American Association for Cancer Research Education Book}, (Karp
  2001):12--16.

\bibitem[Chuang et~al., 2007]{Chuang2007}
Chuang, H.~Y., Lee, E., Liu, Y.~T., Lee, D., and Ideker, T. (2007).
\newblock {Network-based classification of breast cancer metastasis}.
\newblock {\em Molecular System Biology}, 3(140):140.

\bibitem[Danaher et~al., 2014]{Danaher2014}
Danaher, P., Wang, P., and Witten, D.~M. (2014).
\newblock {The joint graphical lasso for inverse covariance estimation across
  multiple classes}.
\newblock {\em Journal of the Royal Statistical Society: Series B (Statistical
  Methodology)}, 76(2):373--397.

\bibitem[Datta, 2001]{Datta2001}
Datta, S. (2001).
\newblock {Exploring relationships in gene expressions: a partial least squares
  approach}.
\newblock {\em Gene Expression}, 9(6):249--255.

\bibitem[de~la Fuente, 2010]{DelaFuente2010}
de~la Fuente, A. (2010).
\newblock {From 'differential expression' to 'differential networking' -
  identification of dysfunctional regulatory networks in diseases}.
\newblock {\em Trends in Genetics}, 26(7):326--333.

\bibitem[Deleforge et~al., 2015a]{DFBH15}
Deleforge, A., Forbes, F., Ba, S., and Horaud, R. (2015a).
\newblock Hyper-spectral image analysis with partially-latent regression and
  spatial markov dependencies.
\newblock {\em IEEE journal of selected topics in signal processing},
  9(6):1037--1048.

\bibitem[Deleforge et~al., 2015b]{DFH14}
Deleforge, A., Forbes, F., and Horaud, R. (2015b).
\newblock High-dimensional regression with gaussian mixtures and
  partially-latent response variables.
\newblock {\em Statistics and Computing}, 25(5):893--911.

\bibitem[Dempster et~al., 1977]{Dempster}
Dempster, A., Laird, N.~M., and Rubin, D.~B. (1977).
\newblock {Maximum likelihood from incomplete data via the EM algorithm.}
\newblock {\em Journal of the Royal Statistical Society. Series B}, Vol.
  39(1):1--38.

\bibitem[Devijver, 2015]{D15}
Devijver, E. (2015).
\newblock Finite mixture regression: a sparse variable selection by model
  selection for clustering.
\newblock {\em Electronic Journal of Statistics}, 9:2642--2674.

\bibitem[Devijver, 2016]{Devijver}
Devijver, E. (2016).
\newblock Model-based clustering for high-dimensional data. {A}pplication to
  functional data.
\newblock {\em Advances in Data Analysis and Classification}, in press.

\bibitem[Devijver and Gallopin, 2017]{DG16}
Devijver, E. and Gallopin, M. (2017).
\newblock Block-diagonal covariance selection for high-dimensional gaussian
  graphical models.
\newblock {\em Journal of the American Statistical Association}, in press.

\bibitem[Dobie et~al., 2001]{Dobie2001}
Dobie, K.~W., Kennedy, C.~D., Velasco, V.~M., McGrath, T.~L., Weko, J.,
  Patterson, R.~W., and Karpen, G.~H. (2001).
\newblock {Identification of chromosome inheritance modifiers in Drosophila
  melanogaster.}
\newblock {\em Genetics}, 157(4):1623--1637.

\bibitem[Dong et~al., 2015]{DYZ15}
Dong, Y., Yu, Z., and Zhu, L. (2015).
\newblock Robust inverse regression for dimension reduction.
\newblock {\em Journal of Multivariate Analysis}, 134:71--81.

\bibitem[Eisen et~al., 1998]{Eisen1998}
Eisen, M.~B., Spellman, P.~T., Brown, P.~O., and Botstein, D. (1998).
\newblock {Cluster analysis and display of genome-wide expression patterns.}
\newblock {\em Proceedings of the National Academy of Sciences},
  95(25):14863--8.

\bibitem[Friedman, 1991]{F91}
Friedman, J. (1991).
\newblock Multivariate adaptive regression splines (with discussion).
\newblock {\em The Annals of Statistics}, 19(1):1--141.

\bibitem[Friedman et~al., 2008]{Friedman2008}
Friedman, J., Hastie, T., and Tibshirani, R. (2008).
\newblock {Sparse inverse covariance estimation with the graphical lasso}.
\newblock {\em Biostatistics}, 9(3):432--441.

\bibitem[Friguet et~al., 2009]{friguet09}
Friguet, C., Kloareg, M., and Causeur, D. (2009).
\newblock A factor model approach to multiple testing under dependence.
\newblock {\em Journal of the American Statistical Association},
  104:488:1406--1415.

\bibitem[Fu, 2013]{Fu2013}
Fu, Y. (2013).
\newblock {Complex networks and simple models in biology}.
\newblock {\em Journal of the Royal Society, Interface}, 2(5):419--30.

\bibitem[Gneiting, 2011]{Gneiting2011}
Gneiting, T. (2011).
\newblock {Making and Evaluating Point Forecasts}.
\newblock {\em Journal of the American Statistical Association},
  106:494:746--762.

\bibitem[Golub, 1999]{Golub1999}
Golub, T.~R. (1999).
\newblock Molecular classification of cancer: class discovery and class
  prediction by gene expression monitoring.
\newblock {\em Science}, 286(5439):531--537.

\bibitem[Hastie et~al., 2010]{HTBbook}
Hastie, T., Tibshirani, R., and Friedman, J. (2010).
\newblock {\em The elements of statistical learning}.
\newblock Springer.

\bibitem[Homyk and Sheppard, 1977]{Homyk1977}
Homyk, T. and Sheppard, D.~E. (1977).
\newblock {Behavioral mutants of Drosophila melanogaster.}
\newblock {\em Genetics}, 87:95--104.

\bibitem[Huang et~al., 2015]{Huang2015}
Huang, W., Carbone, M.~A., Magwire, M.~M., Peiffer, J.~A., Lyman, R.~F., Stone,
  E.~A., Anholt, R. R.~H., and Mackay, T. F.~C. (2015).
\newblock {Genetic basis of transcriptome diversity in Drosophila
  melanogaster.}
\newblock {\em Proceedings of the National Academy of Sciences},
  112(44):E6010--9.

\bibitem[Jacob et~al., 2012]{Jacob2012}
Jacob, L., Neuvial, P., and Dudoit, S. (2012).
\newblock {More power via graph-structured tests for differential expression of
  gene networks.}
\newblock {\em The Annals of Applied Statistics}, 6(2):561--600.

\bibitem[Kraemer et~al., 2008]{KBT08}
Kraemer, N., Boulsteix, A.-L., and Tutz, G. (2008).
\newblock Penalized partial least squares with applications to {B}-spline
  transformations and functional data.
\newblock {\em Chemometrics and ntelligent Laboratory Systems}, 94:60--69.

\bibitem[Kuhn, 2008]{Kuhn2008}
Kuhn, M. (2008).
\newblock {Building Predictive Models in R Using the caret Package}.
\newblock {\em Journal Of Statistical Software}, 28(5):1--26.

\bibitem[{Le Cao} et~al., 2008]{LeCao2008}
{Le Cao}, K.-A., Rossow, D., Robert-Grani{\'{e}}, C., and Besse, P. (2008).
\newblock {A Sparse PLS for Variable Selection when Integrating Omics data}.
\newblock {\em Statistical Applications in Genetics and Molecular Biology},
  7(1):pp. 35.

\bibitem[Leek and Storey, 2007]{leekStorey07}
Leek, J.~T. and Storey, J. (2007).
\newblock Capturing heterogeneity in gene expression studies by surrogate
  variable analysis.
\newblock {\em PLoS Genetics}, 3(9):e161.

\bibitem[Leek and Storey, 2008]{leekStorey08}
Leek, J.~T. and Storey, J. (2008).
\newblock A general framework for multiple testing dependence.
\newblock {\em Proceedings of the National Academy of Sciences},
  105:18718--18723.

\bibitem[Letham et~al., 2015]{Letham2015}
Letham, B., Rudin, C., McCormick, T.~H., and Madigan, D. (2015).
\newblock {Interpretable classifiers using rules and bayesian analysis:
  Building a better stroke prediction model}.
\newblock {\em The Annals of Applied Statistics}, 9(3):1350--1371.

\bibitem[Li, 1991]{L91}
Li, K. (1991).
\newblock Sliced inverse regression for dimension reduction.
\newblock {\em Journal of the American Statistical Association},
  86(414):316--327.

\bibitem[Lloyd et~al., 2000]{Lloyd2000}
Lloyd, T.~E., Verstreken, P., Ostrin, E.~J., Phillippi, A., Lichtarge, O., and
  Bellen, H.~J. (2000).
\newblock {A genome-wide search for synaptic vesicle cycle proteins in
  Drosophila.}
\newblock {\em Neuron}, 26(1):45--50.

\bibitem[Mach et~al., 2013]{Mach2013}
Mach, N., Gao, Y., Lemonnier, G., Lecardonnel, J., Oswald, I.~P.,
  Estell{\'{e}}, J., and Rogel-Gaillard, C. (2013).
\newblock {The peripheral blood transcriptome reflects variations in immunity
  traits in swine: towards the identification of biomarkers.}
\newblock {\em BMC genomics}, 14:894.

\bibitem[Mackay et~al., 2012]{Mackay2012}
Mackay, T. F.~C., Richards, S., Stone, E.~a., Barbadilla, A., Ayroles, J.~F.,
  Zhu, D., Casillas, S., Han, Y., Magwire, M.~M., Cridland, J.~M., Richardson,
  M.~F., Anholt, R. R.~H., Barr{\'{o}}n, M., Bess, C., Blankenburg, K.~P.,
  Carbone, M.~A., Castellano, D., Chaboub, L., Duncan, L., Harris, Z., Javaid,
  M., Jayaseelan, J.~C., Jhangiani, S.~N., Jordan, K.~W., Lara, F., Lawrence,
  F., Lee, S.~L., Librado, P., Linheiro, R.~S., Lyman, R.~F., Mackey, A.~J.,
  Munidasa, M., Muzny, D.~M., Nazareth, L., Newsham, I., Perales, L., Pu,
  L.-L., Qu, C., R{\`{a}}mia, M., Reid, J.~G., Rollmann, S.~M., Rozas, J.,
  Saada, N., Turlapati, L., Worley, K.~C., Wu, Y.-Q., Yamamoto, A., Zhu, Y.,
  Bergman, C.~M., Thornton, K.~R., Mittelman, D., and Gibbs, R.~a. (2012).
\newblock {The Drosophila melanogaster Genetic Reference Panel.}
\newblock {\em Nature}, 482(7384):173--8.

\bibitem[Manwani and Sastry, 2015]{MS15}
Manwani, N. and Sastry, P. (2015).
\newblock K-plane regression.
\newblock {\em Information Sciences}, 292:39 -- 56.

\bibitem[Neum{\"{u}}ller et~al., 2011]{Neumuller2011}
Neum{\"{u}}ller, R.~A., Richter, C., Fischer, A., Novatchkova, M.,
  Neum{\"{u}}ller, K.~G., and Knoblich, J.~A. (2011).
\newblock {Genome-Wide Analysis of Self-Renewal in Drosophila Neural Stem Cells
  by Transgenic RNAi.}
\newblock {\em Cell Stem Cell}, 8(5):580--593.

\bibitem[Nguyen and Rocke, 2002]{Nguyen2002}
Nguyen, D.~V. and Rocke, D.~M. (2002).
\newblock {Tumor classification by partial least squares using microarray gene
  expression data.}
\newblock {\em Bioinformatics}, 18(1):39--50.

\bibitem[Perthame et~al., 2016a]{PFD16}
Perthame, E., Forbes, F., and Deleforge, A. (2016a).
\newblock {Inverse regression approach to robust non-linear high-to-low
  dimensional mapping}.
\newblock preprint hal-01347455.

\bibitem[Perthame et~al., 2016b]{perthame15}
Perthame, E., Friguet, C., and Causeur, D. (2016b).
\newblock Stability of feature selection in classification issues for
  high-dimensional correlated data.
\newblock {\em Statistics and Computing}, 26(4):783--796.

\bibitem[Rhodes and Chinnaiyan, 2005]{Rhodes2005}
Rhodes, D.~R. and Chinnaiyan, A.~M. (2005).
\newblock {Integrative analysis of the cancer transcriptome}.
\newblock {\em Nature Genetics}, 37 Suppl:0--7.

\bibitem[Roberts, 2006]{Roberts2006}
Roberts, D.~B. (2006).
\newblock Drosophila melanogaster: the model organism.
\newblock {\em Entomologia Experimentalis et Applicata}, 121(2):93--103.

\bibitem[Savage, 1971]{Savage}
Savage, L.~J. (1971).
\newblock Elicitation of personal probabilities and expectations.
\newblock {\em Journal of the American Statistical Association},
  66(336):783--801.

\bibitem[Schupbach and Wieschaus, 1991]{Schupbach1991}
Schupbach, T. and Wieschaus, E. (1991).
\newblock {Female sterile mutations on the second chromosome of Drosophila
  melanogaster.}
\newblock {\em Genetics}, 129:1119--1136.

\bibitem[Smyth, 2004]{Smyth2004}
Smyth, G.~K. (2004).
\newblock {Linear Models and Empirical Bayes Methods for Assessing Differential
  Expression in Microarray Experiments Linear Models and Empirical Bayes
  Methods for Assessing Differential Expression in Microarray Experiments}.
\newblock {\em Statistical Applications in Genetics and Molecular Biology},
  3(1):1--26.

\bibitem[Suzuki et~al., 2014]{Suzuki2014}
Suzuki, S., Horinouchi, T., and Furusawa, C. (2014).
\newblock {Prediction of antibiotic resistance by gene expression profiles}.
\newblock {\em Nature Communications}, 5:5792.

\bibitem[Swarup et~al., 2013]{Swarup2013}
Swarup, S., Huang, W., Mackay, T. F.~C., and Anholt, R. R.~H. (2013).
\newblock {Analysis of natural variation reveals neurogenetic networks for
  Drosophila olfactory behavior.}
\newblock {\em Proceedings of the National Academy of Sciences of the United
  States of America}, 110(3):1017--22.

\bibitem[Tipping, 2001]{T01}
Tipping, M.~E. (2001).
\newblock Sparse bayesian learning and the relevance vector machine.
\newblock {\em The Journal of Machine Learning Research}, 1:211--244.

\bibitem[Torres-Garc{\'{i}}a et~al., 2009]{Torres-Garcia2009}
Torres-Garc{\'{i}}a, W., Zhang, W., Runger, G.~C., Johnson, R.~H., and Meldrum,
  D.~R. (2009).
\newblock {Integrative analysis of transcriptomic and proteomic data of
  Desulfovibrio vulgaris: A non-linear model to predict abundance of undetected
  proteins}.
\newblock {\em Bioinformatics}, 25(15):1905--1914.

\bibitem[Valc{\'{a}}rcel et~al., 2014]{Valcarcel2014}
Valc{\'{a}}rcel, B., Ebbels, T. M.~D., Kangas, A.~J., Soininen, P., Elliot, P.,
  Ala-Korpela, M., J{\"{a}}rvelin, M.-R., and de~Iorio, M. (2014).
\newblock {Genome metabolome integrated network analysis to uncover connections
  between genetic variants and complex traits: an application to obesity.}
\newblock {\em Journal of the Royal Society, Interface}, 11(94):20130908.

\bibitem[Valc{\'{a}}rcel et~al., 2011]{Valcarcel2011}
Valc{\'{a}}rcel, B., Wurtz, P., al~Basatena, N. K.~S., Tukiainen, T., Kangas,
  A.~J., Soininen, P., J{\"{a}}rvelin, M.-R., Ala-Korpela, M., Ebbels, T.~M.,
  and de~Iorio, M. (2011).
\newblock {A differential network approach to exploring differences between
  biological states: An application to prediabetes}.
\newblock {\em PLoS ONE}, 6(9).

\bibitem[Vapnik, 1998]{V98}
Vapnik, V. (1998).
\newblock {\em Statistical Learning Theory}.
\newblock Wiley, New York.

\bibitem[Wishart et~al., 2012]{Wishart2012}
Wishart, T.~M., Rooney, T.~M., Lamont, D.~J., Wright, A.~K., Morton, A.~J.,
  Jackson, M., Freeman, M.~R., and Gillingwater, T.~H. (2012).
\newblock {Combining comparative proteomics and molecular genetics uncovers
  regulators of synaptic and axonal stability and degeneration in vivo.}
\newblock {\em PLoS Genetics}, 8(8):e1002936.

\bibitem[Wold et~al., 1989]{WKS89}
Wold, S., Kettaneh-Wold, N., and Skagerberg, B. (1989).
\newblock Nonlinear pls modeling.
\newblock {\em Chemometrics and Intelligent Laboratory Systems}, 7:53--65.

\bibitem[Wu, 2008]{W08}
Wu, H. (2008).
\newblock Kernel sliced inverse regression with applications to classification.
\newblock {\em Journal of Computational and Graphical Statistics},
  17(3):590--610.

\bibitem[Yang et~al., 2011]{Yang2011}
Yang, B., Bassols, A., Saco, Y., and P{\'{e}}rez-Enciso, M. (2011).
\newblock {Association between plasma metabolites and gene expression profiles
  in five porcine endocrine tissues.}
\newblock {\em Genetics, Selection, Evolution}, 43(1):28.

\bibitem[Yugi et~al., 2016]{Yugi2016}
Yugi, K., Kubota, H., Hatano, A., and Kuroda, S. (2016).
\newblock {Trans-Omics : How To Reconstruct Biochemical Networks Across
  Multiple 'Omic' Layers}.
\newblock {\em Trends in Biotechnology}, 34(4):276--290.

\bibitem[Zhong et~al., 2005]{ZZ05}
Zhong, W., Zeng, P., Ma, P., Liu, J., and Zhu, Y. (2005).
\newblock Rsir: regularized sliced inverse regression for motif discovery.
\newblock {\em Bioinformatics}, 21(22):4169--4175.

\end{thebibliography}

\appendix
\section{Appendix}

This appendix contains the details of the updates of parameters estimations of our algorithm. 

\subsection{Details on the E-step of algorithm}

Individual probabilities are updated by, for the iteration $(\text{ite})$,
\begin{align*}
 r^{(\text{ite})}_{i,k} &= \frac{\pi_k^{(\text{ite}-1)}\Prob(\Yv_i,\Xv_i \vert Z_i=k ; \thetab^{(\text{ite}-1)})}{\sum_{j=1}^K \pi_j^{(\text{ite}-1)}\Prob(\Yv_i,\Xv_i \vert Z_i=j ;\thetab^{(\text{ite}-1)})}.
\end{align*}
These probabilities can be expressed in function of all marginals using Bayes formula. 
\subsection{Details on the M-step of algorithm}

\paragraph{{\bf GMM-like estimators}}
 $\pi_k$, $\cb_k$ and $\Gammab_k$ are estimated by GMM-like estimators as in \cite{DFH14}
 
\begin{align*}
 \pi^{(\text{ite})}_k &= \frac{\sum_{i=1}^n  r^{(\text{ite})}_{i,k}}{n} \\
 \cb_k^{(\text{ite})} &= \sum_{i=1}^n \frac{ r^{(\text{ite})}_{i,k}}{ \sum_{j=1}^n r^{(\text{ite})}_{j,k}} \xv_i \\
 \Gammab_k^{(\text{ite})} &= \sum_{i=1}^n \frac{ r^{(\text{ite})}_{i,k}}{ \sum_{j=1}^n r^{(\text{ite})}_{j,k}} (\xv_i -  \cb_k^{(\text{ite})})(\xv_i -  \cb_k^{(\text{ite})})^T
\end{align*}

\paragraph{\bf Regression-like estimators}
 For $k \in \{1,\ldots, K\}$, $\Ab_k$ and $\bb_k$ are estimated by Regression-like estimators as in \cite{DFH14}

\begin{align*}
 \Ab_k^{(\text{ite})} &=  \Yv^{(\text{ite})}_k  (\xv^{(\text{ite})}_k)^T ( \xv^{(\text{ite})}_k (\xv_k^{(\text{ite})})^T)^{-1} \\
 \bb^{(\text{ite})}_k &= \sum_{i=1}^n \frac{ r^{(\text{ite})}_{i,k}}{ \sum_{j=1}^n r^{(\text{ite})}_{j,k}}(\Yv_i -  \Ab^{(\text{ite})}_k \xv_{i,k})
\end{align*}

where $\xv_k^{(\text{ite})}$ and $ \Yv_k^{(\text{ite})}$ 
are the observations reweighted by the cluster weights and are expressed in function of the observed data:
\begin{align*}
 \xv_k^{(\text{ite})} &= \frac{1}{\sqrt{\sum_{i=1}^n  r^{(\text{ite})}_{i,k}}} \left[\sqrt{ r^{(\text{ite})}_{1,k}}( \xv^{(\text{ite})}_{1,k} -  \sum_{i=1}^n \frac{ r^{(\text{ite})}_{i,k}}{\sum_{j=1}^n  r^{(\text{ite})}_{j,k}}  \xv_{i,k}), \ldots, \sqrt{ r^{(\text{ite})}_{N,k}}( \xv^{(\text{ite})}_{n,k} -  \sum_{i=1}^n \frac{ r^{(\text{ite})}_{i,k}}{\sum_{j=1}^n  r^{(\text{ite})}_{j,k}}  \xv_{i,k})\right]\\
 \Yv^{(\text{ite})}_k &= \frac{1}{\sqrt{ \sum_{i=1}^n r^{(\text{ite})}_{i,k}}} \left[\sqrt{ r^{(\text{ite})}_{1,k}} (\Yv_{1} -  \sum_{i=1}^n \frac{ r^{(\text{ite})}_{i,k}}{ \sum_{j=1}^n r^{(\text{ite})}_{j,k}} \Yv_{i}), \ldots, \sqrt{ r^{(\text{ite})}_{n,k}}( \Yv_{n} -  \sum_{i=1}^n \frac{ r^{(\text{ite})}_{i,k}}{ \sum_{j=1}^n r^{(\text{ite})}_{j,k}} \Yv_{i})\right]
\end{align*}

\paragraph{\bf Block-diagonal covariance estimator}
In GLLiM, for $k \in \{1, \ldots, K\}$, $\Sigmab_k$ is estimated as the diagonal of the residual covariance matrix of the regression of $\Yv_i$ on $\Xv_i$ in each cluster $k$ and weighted by the cluster weights $\frac{ r_{i,k}}{ \sum_{i=1}^n r^{(\text{ite})}_{i,k}}$. In this paper, we propose to replace this estimation step by a block-diagonal covariance matrix, and then $\Sigmab_k$ is estimated by a block-diagonal estimator, as explained in Section 2.2.

\end{document}